\documentclass[twocolumn]{aastex62}

\usepackage{graphicx,amsmath}
\usepackage{textcomp}
\usepackage{amssymb}
\usepackage{nccmath}
\usepackage{upgreek}
\usepackage{subfigure}
\usepackage{color}
\usepackage{bm}
\usepackage{soul}
\usepackage{CJK}
\usepackage{ulem}
\hypersetup{linkcolor=blue, citecolor=blue}

\setcitestyle{notesep={ }}
\newcommand{\hi}{H\thinspace{\sc i}}

\newcommand{\ha}{H$\alpha$}
\newcommand{\hb}{H$\beta$}

\newcommand{\Msol}{\hbox{\thinspace $M_{\odot}$}}

\newcommand{\htwo}{H$_2$}

\shorttitle{The gas cycle of local galaxy populations}
\shortauthors{Dou et al.}

\begin{document}

\title{{\bf \Large From haloes to galaxies. III. The gas cycle of local galaxy populations}}

\correspondingauthor{Yingjie Peng}
\email{yjpeng@pku.edu.cn}

\author[0000-0002-6961-6378]{Jing Dou}
\affiliation{Kavli Institute for Astronomy and Astrophysics, Peking University, 5 Yiheyuan Road, Beijing 100871, China}
\affiliation{Department of Astronomy, School of Physics, Peking University, 5 Yiheyuan Road, Beijing 100871, China}

\author{Yingjie Peng}
\affiliation{Kavli Institute for Astronomy and Astrophysics, Peking University, 5 Yiheyuan Road, Beijing 100871, China}

\author[0000-0002-7093-7355]{Alvio Renzini}
\affiliation{INAF - Osservatorio Astronomico di Padova, Vicolo dell'Osservatorio 5, I-35122 Padova, Italy}

\author[0000-0001-6947-5846]{Luis C. Ho}
\affil{Kavli Institute for Astronomy and Astrophysics, Peking University, 5 Yiheyuan Road, Beijing 100871, China}
\affil{Department of Astronomy, School of Physics, Peking University, 5 Yiheyuan Road, Beijing 100871, China}

\author[0000-0002-4803-2381]{Filippo Mannucci}
\affiliation{INAF - Osservatorio Astrofisico di Arcetri, Largo Enrico Fermi 5, I-50125 Firenze, Italy}

\author[0000-0002-3331-9590]{Emanuele Daddi}
\affiliation{CEA, Irfu, DAp, AIM, Universit\'{e} Paris-Saclay, Universit\'{e} de Paris, CNRS, F-91191 Gif-sur-Yvette, France}

\author[0000-0003-0007-2197]{Yu Gao}
\affiliation{Department of Astronomy, Xiamen University, Xiamen, Fujian 361005, China}
\affiliation{Purple Mountain Observatory $\&$ Key Laboratory for Radio Astronomy, Chinese Academy of Sciences, 10 Yuanhua Road, Nanjing 210033, PR China}

\author[0000-0002-4985-3819]{Roberto Maiolino}
\affiliation{Cavendish Laboratory, University of Cambridge, 19 J. J. Thomson Avenue, Cambridge CB3 0HE, UK}
\affiliation{Kavli Institute for Cosmology, University of Cambridge, Madingley Road, Cambridge CB3 0HA, UK}
\affiliation{Department of Physics and Astronomy, University College London, Gower Street, London WC1E 6BT, UK}

\author[0000-0001-6469-1582]{Chengpeng Zhang}
\affiliation{Kavli Institute for Astronomy and Astrophysics, Peking University, 5 Yiheyuan Road, Beijing 100871, China}
\affiliation{Department of Astronomy, School of Physics, Peking University, 5 Yiheyuan Road, Beijing 100871, China}

\author[0000-0002-3890-3729]{Qiusheng Gu}
\affil{School of Astronomy and Space Science, Nanjing University, Nanjing 210093, China.}
\affil{Key Laboratory of Modern Astronomy and Astrophysics (Nanjing University), Ministry of Education, Nanjing 210093, China.}

\author[0000-0003-3010-7661]{Di Li}
\affiliation{CAS Key Laboratory of FAST, National Astronomical Observatories, Chinese Academy of Sciences, Beijing 100012, China}
\affiliation{School of Astronomy and Space Science, University of Chinese Academy of Sciences, Beijing 100049, China}

\author[0000-0002-6423-3597]{Simon J. Lilly}
\affiliation{Department of Physics, ETH Zurich, Wolfgang-Pauli-Strasse 27, CH-8093 Zurich, Switzerland}

\author[0000-0001-5662-8217]{Zhizheng Pan}
\affiliation{Purple Mountain Observatory, Chinese Academy of Sciences, 10 Yuan Hua Road, Nanjing, Jiangsu 210033, China}
\affiliation{School of Astronomy and Space Sciences, University of Science and Technology of China, Hefei, 230026, China}

\author[0000-0003-3564-6437]{Feng Yuan}
\affiliation{Key Laboratory for Research in Galaxies and Cosmology, Shanghai Astronomical Observatory, Chinese Academy of Sciences, 80 Nandan Road, Shanghai 200030, China}

\author[0000-0003-3728-9912]{Xianzhong Zheng}
\affiliation{Purple Mountain Observatory, Chinese Academy of Sciences, 10 Yuan Hua Road, Nanjing, Jiangsu 210033, China}
\affiliation{School of Astronomy and Space Sciences, University of Science and Technology of China, Hefei, 230026, China}

\begin{abstract}

\noindent In Dou et al. (2021), we introduced the Fundamental Formation Relation (FFR), a tight relation between specific SFR (sSFR), \htwo\ star formation efficiency (SFE$_{\rm H_2}$), and the ratio of \htwo\ to stellar mass. Here we show that atomic gas \hi\ does not follow a similar FFR as \htwo. The relation between SFE$_{\rm HI}$ and sSFR shows significant scatter and strong systematic dependence on all of the key galaxy properties that we have explored. The dramatic difference between \hi\ and \htwo\ indicates that different processes (e.g., quenching by different mechanisms) may have very different effects on the \hi\ in different galaxies and hence produce different SFE$_{\rm HI}$-sSFR relations, while the SFE$_{\rm H_2}$-sSFR relation remains unaffected. The facts that SFE$_{\rm H_2}$-sSFR relation is independent of other key galaxy properties, and that sSFR is directly related to the cosmic time and acts as the cosmic clock, make it natural and very simple to study how different galaxy populations (with different properties and undergoing different processes) evolve on the same SFE$_{\rm H_2}$-sSFR $\sim t$ relation. In the gas regulator model (GRM), the evolution of a galaxy on the SFE$_{\rm H_2}$-sSFR($t$) relation is uniquely set by a single mass-loading parameter $\lambda_{\rm net,H_2}$. This simplicity allows us to accurately derive the \htwo\ supply and removal rates of the local galaxy populations with different stellar masses, from star-forming galaxies to the galaxies in the process of being quenched. This combination of FFR and GRM, together with the stellar metallicity requirement, provide a new powerful tool to study galaxy formation and evolution.

\end{abstract}


\section{Introduction} \label{sec:intro}

\defcitealias{2021Dou}{D21}

From the first principle, the global star formation rates (SFRs) of galaxies are determined by their cold gas content ($M_{\rm gas}$) and star formation efficiency (SFE, defined as SFR/$M_{\rm gas}$). The gas content in the galaxy is regulated by the dynamical balance among gas cooling and accretion from the intergalactic medium, gas consumption from star formation, gas outflow driven by feedback, and gas stripping. Gas content and star formation are expected to be primarily controlled by the net gas accretion (i.e. gas inflow minus gas outflow or gas removal), and star formation proceeds in a self-regulated manner. For a given positive net gas accretion rate, a higher SFR consumes gas faster and gas mass will decrease, resulting in a declining SFR; while a lower SFR consumes gas slower and gas mass will increase, resulting in a rising SFR. Quenching plays a critical role in gas cycle and star formation. For instance, star-forming galaxies can be quenched by gas removal, via either external environmental process \citep[e.g.,][]{1972ApJ...176....1G,1999MNRAS.308..947A,2000Sci...288.1617Q} or ejection by feedback \citep[e.g.,][]{2004ApJ...600..580G}. It can also be quenched by strangulation, being broadly interpreted as halting gas inflow via either external enviroment process \citep[e.g.,][]{1980ApJ...237..692L,2000MNRAS.318..703B,2000ApJ...540..113B,2006MNRAS.368....2D} or internal processes such as preventive feedback \citep[e.g.,][]{2006MNRAS.365...11C,2015ARA&A..53...51S,2020Zinger}; or the excess angular momentum of the inflowing gas \citep[e.g.,][]{2020MNRAS.491L..51P,2020MNRAS.495L..42R,2020arXiv200900013S}.
Despite all these plausible quenching mechanisms that may act at different epochs or in different environments, if net gas accretion resumes (i.e. inflow larger than outflow or gas removal), the quenched galaxy may rejuvenate as gas gradually accumulates in the galaxy and continue to form stars.

Therefore, understanding gas cycle and star formation process are central issues in studying galaxy formation and evolution \citep[e.g.,][]{1978MNRAS.183..341W,2006MNRAS.368....2D,2011MNRAS.415.2782V,2012MNRAS.421...98D,2013MNRAS.430.1051C,2014ARA&A..52..415M,2017ApJ...837..150S,Tacconi:2018jb,2018MNRAS.477.2716K,2019ApJ...885L..14P,2020Walter}. By analyzing the molecular gas content in the local galaxy populations, we show in \citeauthor{2021Dou} (2021; hereafter D21) that the specific relations (i.e. $\mu$-sSFR, SFE-sSFR, and SFE-$\mu$, where $\mu$ is the \htwo\ gas mass to stellar mass ratio) are much tighter than the corresponding absolute ones (i.e. $M_{\rm H_2}$-SFR, $M_{\rm H_2}$-$M_*$, and SFR-$M_*$). In particular, the scatter of the $\mu$-sSFR and SFE–sSFR relations can be entirely explained by the measurement errors, which implies that the intrinsic scatter of the $\mu$-sSFR and SFE–sSFR relations is very small. This also suggests that there is little room to further reduce the scatter of these specific relations by including any systematic dependence on other galaxy properties. Indeed, as shown in \citetalias{2021Dou}, the specific relations are independent of all other key galaxy properties that we have explored, including stellar mass, structure, environment and metallicity. This suggests some more universal or important physical connection between these quantities.

The small intrinsic scatters of the specific relations also require them to have only one single sequence holding from star-bursting galaxies to galaxies in the process of being quenched, as indeed observed and shown in \citetalias{2021Dou}. On the contrary, the absolute relations contain two structures (i.e. a star-forming sequence and a passive cloud), which contribute to increase their overall scatters. The large scatters of the absolute relations are also due to their strong systematic dependence on other galaxies properties. Therefore, we have proposed the sSFR-$\mu$-SFE relation as the Fundamental Formation Relation (FFR), which governs the star formation and quenching processes, and provides a simple framework to study galaxy evolution.

The three quantities in FFR are linked by equation sSFR = $\mu \times$SFE. sSFR, SFE, and $\mu$ are not only normalized quantities of $M_*$, SFR, and $M_{\rm H_2}$, but are also primary parameters in galaxy formation and evolution with specific physical meanings. As discussed in Section 5.1 in \citetalias{2021Dou}, 1/sSFR is the e-folding timescale of the growth of the stellar mass. The evolution of sSFR is primarily driven by the dark matter halo accretion history \citep{Peng:2014hn}. SFE (or the gas depletion timescale $\tau$) is related to a galaxy's dynamical time. Some fraction of molecular gas is turned into stars per galactic orbital time under the gravitational instability of the cold gas on the disk, and the exact fraction depends on detailed feedback physics \citep{1997ApJ...480..235E,1997Silk,1998Kennicutt,2010Genzel}. $\mu$ is mainly a measurement of the gravitational instability parameter $Q$ for the gas disk \citep[e.g.,][]{2002Wong}. Therefore, the equation sSFR = $\mu \times$SFE suggests that the star formation level, in terms of sSFR, is determined by the combination of gas instability and the galactic dynamic timescale of the disk. It is also interesting to note that 1/SFE and 1/sSFR are both timescales, one for gas and one for stars. 1/SFE is termed as “galactic clock” in \citet{2020Tacconi} and 1/sSFR is called as “cosmic clock” in \citet{Peng:2010gn}.

In parallel, the gas regulator model \citep[e.g.,][]{2010ApJ...718.1001B,Lilly:2013ko,Peng:2014hn,2014MNRAS.444.2071D,2019MNRAS.487..456B} has been proposed as a simple toy model to describe the interplay between gas accretion, star formation and outflow. Despite its simplicity, the model has been quite successful in interpreting a variety of observations of star forming, secularly-evolving galaxies, including the mass-metallicity relation \citep[e.g.,][]{1979A&A....80..155L,2004ApJ...613..898T}, the fundamental metallicity relation \citep[e.g.,][]{2010MNRAS.408.2115M}, metallicity gradients \citep[e.g.,][for a review]{2019A&ARv..27....3M} and quenching mechanism from the stellar metallicity difference between star-forming and passive galaxies \citep{Peng:2015bq,2020MNRAS.491.5406T,2020MNRAS.tmp.3344T}.

In this paper, we first explore whether the \hi\ gas follows a similar FFR as \htwo. Then we explore how to use the FFR of the molecular gas, in combination with the gas regulator model, to study galaxy evolution.  

\section{Sample} \label{sec:sample}

The main sample analyzed in this paper is the same xCOLD GASS \citep{2017Saintonge} sample that we have used in \citetalias{2021Dou}. Briefly, it was assembled through CO (1-0) observations on the IRAM 30m single-dish telescope. The xCOLD GASS survey aims to uniformly span the SFR-stellar mass ($M_*$) plane down to $M_* \sim 10^{9}\Msol$ in the nearby Universe ($0.01<z<0.05$), and hence is not biased towards star-forming galaxies or passive galaxies. This makes it an ideal sample to study both star formation and quenching processes in the local galaxies. Galaxies are selected in xCOLD GASS to produce a roughly flat distribution in $M_*$, which is different from the mass distribution of the parent SDSS sample. This mass bias can be corrected by a statistical weight \citep{2010MNRAS.403..683C}.

The \htwo~gas mass of each individual galaxy is derived from CO luminosity via the CO-to-\htwo~conversion factor ($\alpha_{\rm CO}$) using the calibrations in \citet{2017Accurso} as recommended by xCOLD GASS. We have also tested alternative conversion factors, including constant $\alpha_{\rm CO}$ or metallicity-dependent-only $\alpha_{\rm CO}$ \citep[e.g.,][]{2012ApJ...746...69G}. The results are very similar to these presented in the paper, hence our results are not sensitive to the choice of conversion factor. The stellar masses are retrieved from the SDSS DR7 MPA-JHU catalog \citep{2007Salim}. SFRs in xCOLD GASS are calculated using the combination of MIR and UV from WISE and GALEX survey, respectively, as described in \citet{2017Janowiecki}. All stellar masses and SFRs are converted to a Chabrier initial mass function \citep{2003PASP..115..763C}. The final sample used in our analysis contains 330 galaxies with reliable CO (1-0), SFR and stellar mass measurements.

Other auxiliary data used in our analysis includes r-band effective radius (R$_{50}$), which is obtained from the SDSS DR7 \citep{Abazajian:2009ef} official database. The mass-weighted bulge-to-total ratios (B/T) are taken from \citet{Simard2011} and \citet{2014ApJS..210....3M}, where a pure exponential disk and a de Vaucouleurs bulge are used. The classifications of the central and satellite galaxies are retrieved from the SDSS DR7 \citet{Yang:2007} group catalogue. Central galaxies are defined to be both the most massive and the most luminous (r-band) galaxy within a given group. Other galaxies in the group are defined as satellites. The gas-phase metallicities 12 + log O/H were measured from the emission line ratios derived by \citet{2004ApJ...613..898T}. Only galaxies with lines of \hb, \ha, and [\ion{N}{2}]~$\lambda 6584$ detected at greater than 5 $\sigma$ have metallicity measurements, and galaxies with weaker or no emission lines (i.e. most of the quenching or quenched galaxies) are not included in the metallicity-related analysis. We have also tested other independent measurements of metallicity such as the N2 and O3N2 calibrations and find very small changes to the results presented in this paper.

The FFR proposed in \citetalias{2021Dou} refers to the molecular gas \htwo. In this work, we also wish to explore if the \hi\ follows a similar FFR as the \htwo\ gas, by including in our analysis the \hi\ gas measurements from the extended GALEX Arecibo SDSS Survey (xGASS). xGASS provides a census of the \hi\ gas content of 1179 galaxies using the Arecibo telescope \citep{2018Catinella}. They are selected only by redshift ($0.01<z<0.05$) and stellar mass ($10^{9}\Msol<M_*<10^{11.5}\Msol$) from the intersected area of the SDSS DR7 spectroscopic survey, the GALEX Medium Imaging Survey \citep{Martin:2005} and projected ALFALFA footprints \citep{Haynes:2011en}. Similar to xCOLD GASS, the xGASS survey is composed of GASS and GASS-low (low mass extension of GASS). To optimize the survey efficiency, galaxies with reliable \hi\ detections already available from the 40\% ALFALFA catalog or the Cornell \hi\ digital archive \citep{Springob:2005vt}, were not observed again. The rest galaxies were observed with the Arecibo telescope until the \hi\ line was detected, or a limit of a few percent in $M_{\rm HI}/M_*$ was reached. This limit of $M_{\rm HI}/M_*$ is 2\% for galaxies with log $M_*>$ 9.7, and a constant gas mass limit of log $M_{\rm HI}$ = 8 for galaxies with lower stellar masses (which corresponds to a varying limit of $M_{\rm HI}/M_*$ from 2\% at log $M_* \sim$ 9.7, 4\% at log $M_* \sim$ 9.4 to 10\% at log $M_* \sim$ 9). In our analysis, we only include reliable \hi~detections such that the \hi~line is detected and not confused by close companions. After this selection, there are 662 galaxies in our \hi~sample. Each galaxy is weighted by a correction factor to account for selection effects in stellar mass, described in \citet{2018Catinella}. The selection effects were discussed in detail in the Appendix in \citetalias{2021Dou}. In addition, we show the \hi~detection ratio as a function of SFR and $M_*$ (Figure A1), and of sSFR (Figure A2), analogous to Figure A1 and A2 in \citetalias{2021Dou} for the \htwo\ detection ratio.

\section{Results}

\subsection{Molecular and atomic hydrogen gas} \label{sec:gas}

\begin{figure*}[htbp]
    \begin{center}
       \includegraphics[width=180mm]{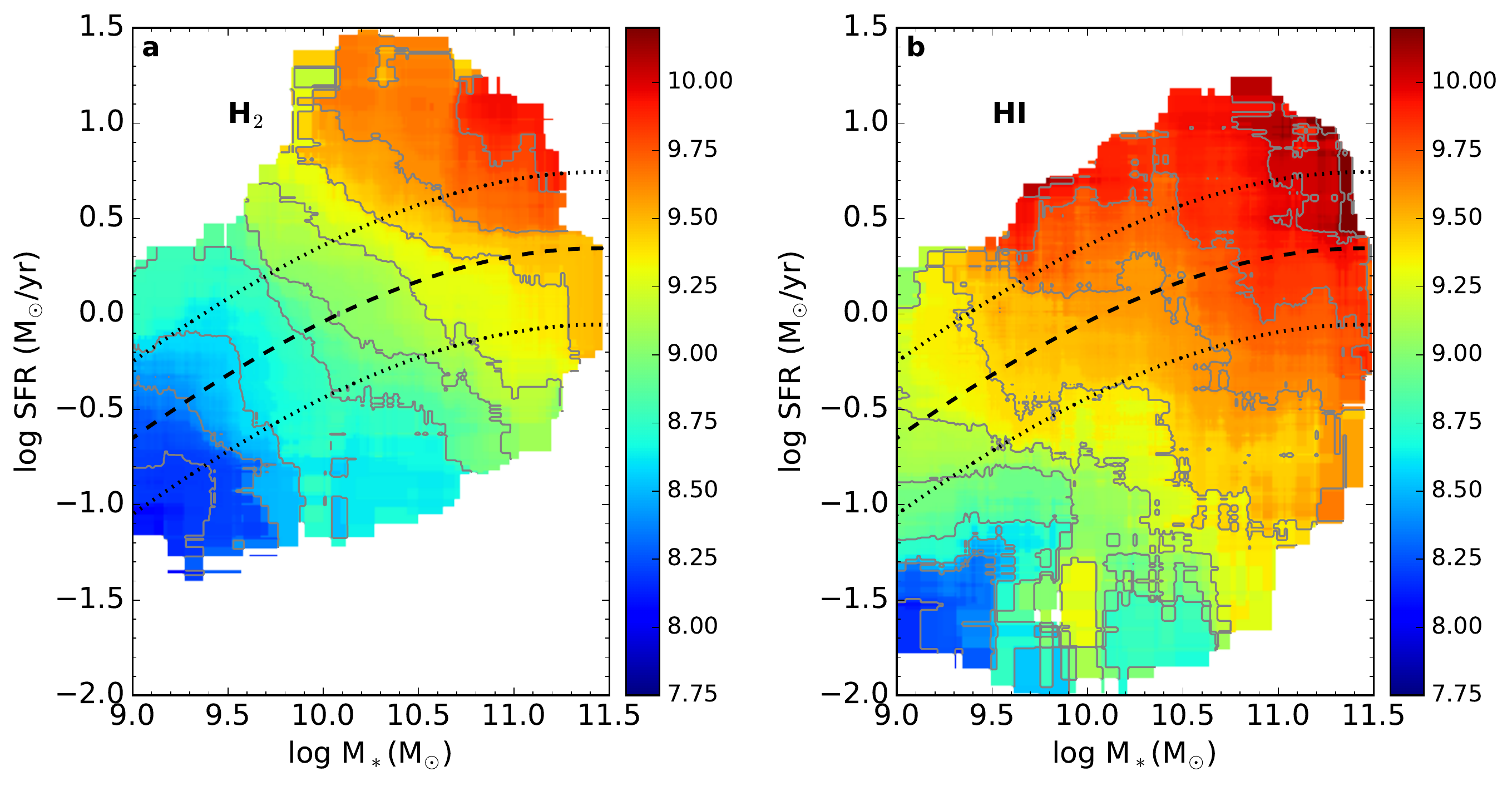}
    \end{center}
\caption{The average molecular gas mass (left panel) and average atomic gas mass (right panel) in the SFR-$M_*$ plane. The average values are obtained by using a moving box of size 0.6 dex in SFR and 0.6 dex in $M_*$, with the requirement of minimum of 3 galaxies in each box. In both panels, the dashed line indicates the position of the star-forming main sequence defined in \citet{2016MNRAS.462.1749S}. The dotted lines indicate $\pm$0.4 dex scatter around the main sequence.}
 \label{h2hi_2d}
\end{figure*}

Figure \ref{h2hi_2d} shows the average \htwo~molecular gas mass (left panel) and average \hi~gas mass (right panel) in the SFR-$M_*$ plane. In both panels, the dashed line indicates the position of the star-forming main sequence (MS) defined in \citet{2016MNRAS.462.1749S}. The dotted lines indicate $\pm$0.4 dex scatter around the main sequence. The lower dotted line is the approximate divide between star-forming and quiescent galaxies. The upper dotted line indicates the approximate divide between normal star-forming galaxies and galaxies with elevated SFR (starburst galaxies). It is evident that the color-coded data distribution in panel a and b are different. This is because panel a is produced by galaxies with \htwo~detections regardless of their \hi~detection status, while panel b is produced by galaxies with \hi~detections regardless of their \htwo~detection status. These two samples share many common galaxies that have detections in both \hi~and \htwo, but are not the same due to that some galaxies have only \hi~detections/observations while some have only \htwo~detections/observations. We discuss below the reasons that cause this difference in data distribution and the impact to the results.

Comparing the data distribution in panel a and b, the small extra extension to the star-bursting regime in \htwo~at log $M_* \sim$ 10 - 11 is due to the xCOLD GASS survey containing some extra star-bursting galaxies that have not been included in the xGASS survey. In other words, this is not because these star-bursting galaxies have \htwo~detections but no \hi~detections, but simply because they have not been observed in \hi. Meanwhile, the fact that many starbursts/LIRGs have strong 21-cm continuum in the star-bursting regions results in strong \hi~absorption and hence obvious \hi~emission may not be easily detected by single-dish telescopes \citep[e.g.,][]{2001van}. The more obvious difference between panel a and b is in the low SFR regime, where only \hi~data is present. First, this difference here is not due to different samples as in the star-bursting regime discussed above. These galaxy samples are identical, i.e. they have been observed in both \hi~and \htwo. Second, the observation limit in \htwo~is similar to \hi, both go down to a gas fraction of about 2\%. This suggests that some galaxies in the low SFR regime are genuine \htwo~poor (hence not detected in CO) but \hi~rich (hence detected in \hi). Indeed, as discussed in \citet{2019ApJ...884L..52Z,2021zhang}, during the quenching of the massive central disk galaxies, their \htwo~gas mass drops rapidly by more than 1 dex from MS to the lowest observed sSFR, but their \hi~gas mass remains surprisingly constant, i.e. the \hi/\htwo~ratio rapidly increases with decreasing sSFR. If we select galaxies that have detections in both \hi~and \htwo, the results in panel a and b in the overlapped region of \hi~and \htwo~distributions on the SFR-$M_*$ plane remain the same.

Comparing panels a and b, \htwo~gas mass and \hi~gas mass follow similar trend of distribution on the SFR-$M_*$ plane. Both of them increase with increasing stellar mass and SFR. At a given stellar mass and SFR, on average, the \hi~gas mass is higher than the \htwo~gas mass, especially for low-mass galaxies that are \hi\ dominated. This indicates the SFE of \htwo\ gas (SFE$_{\rm H_2}$) is higher than that of the \hi\ gas (SFE$_{\rm HI}$). Or equivalently, the gas depletion timescale $\tau_{\rm dep}$ (=1/SFE) of \htwo\ gas ($\tau_{\rm dep,H_2}$) is shorter than that of the \hi\ gas ($\tau_{\rm dep,HI}$). It should be noted that SFE$_{\rm HI}$ can be reformed as SFE$_{\rm HI}$ = SFE$_{\rm H_2} \times M_{\rm H_2}/M_{\rm HI}$. Hence SFE$_{\rm HI}$ is simply the product of SFE$_{\rm H_2}$ and the \htwo~to \hi~gas mass ratio.

The typical timescale of \htwo\ formation in equilibrium is $\sim 10^7$ yr \citep[e.g.,][]{2007ApJ...654..273G,2007A&A...461..205L}, which is much shorter than $\tau_{\rm dep,H_2}$ of typical star-forming galaxies. This hence suggests that a significant fraction of \hi\ gas will not be able to form \htwo, in particular in low-mass galaxies, probably due to the low surface density or lack of dust. \htwo\ formation efficiency is also expected to depend on the gas-phase metallicity and/or the gas pressure \citep[e.g.,][]{2006Blitz,2009Krumholz}. In addition, \htwo\ can also be destroyed by star formation feedback \citep{2020MNRAS.496.4606M}.

\subsection{SFE-sSFR relation} \label{sec:sfe-ssfr}

\begin{figure*}[htbp]

    \includegraphics[width=1\linewidth]{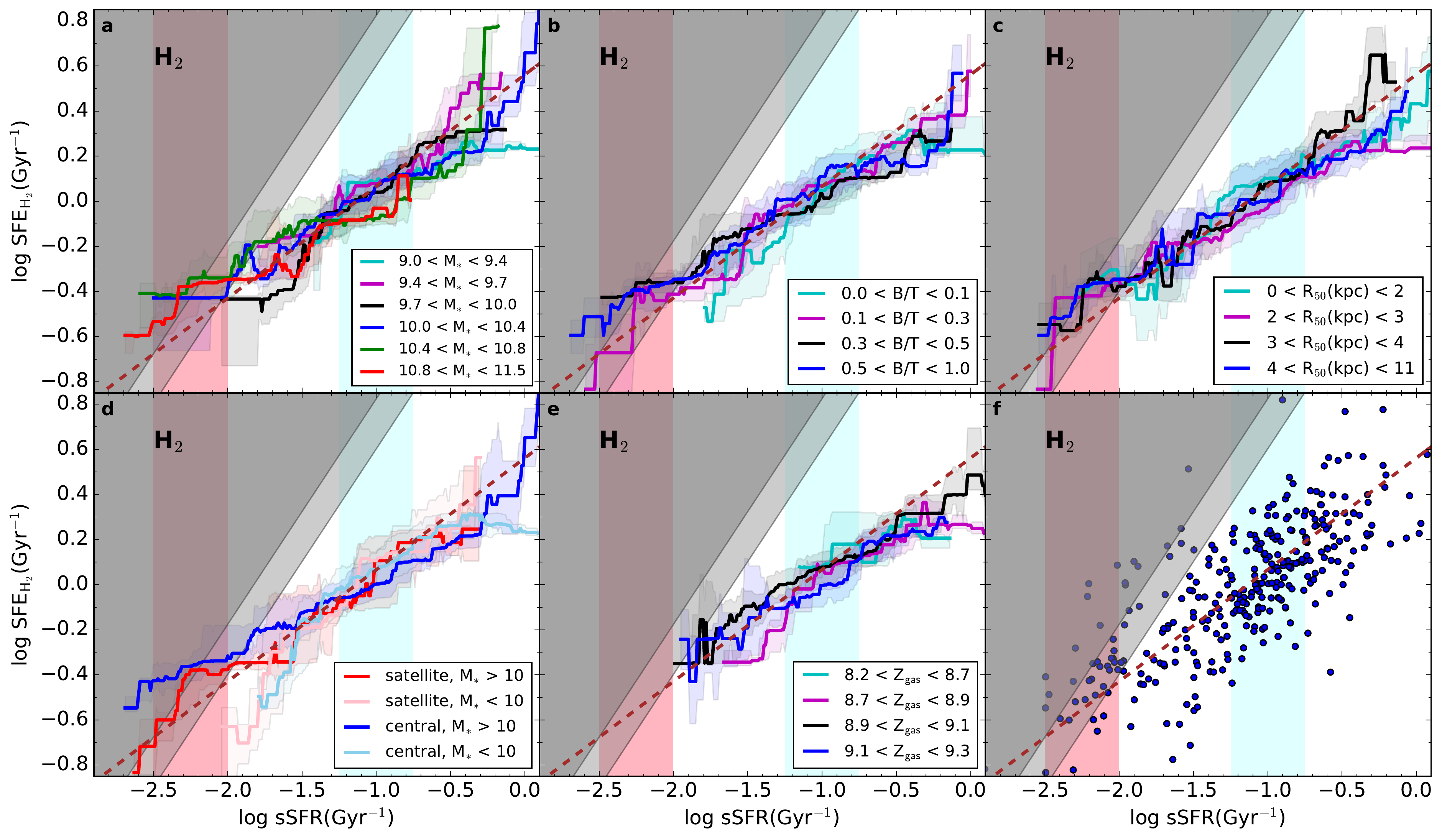}
    \includegraphics[width=1\linewidth]{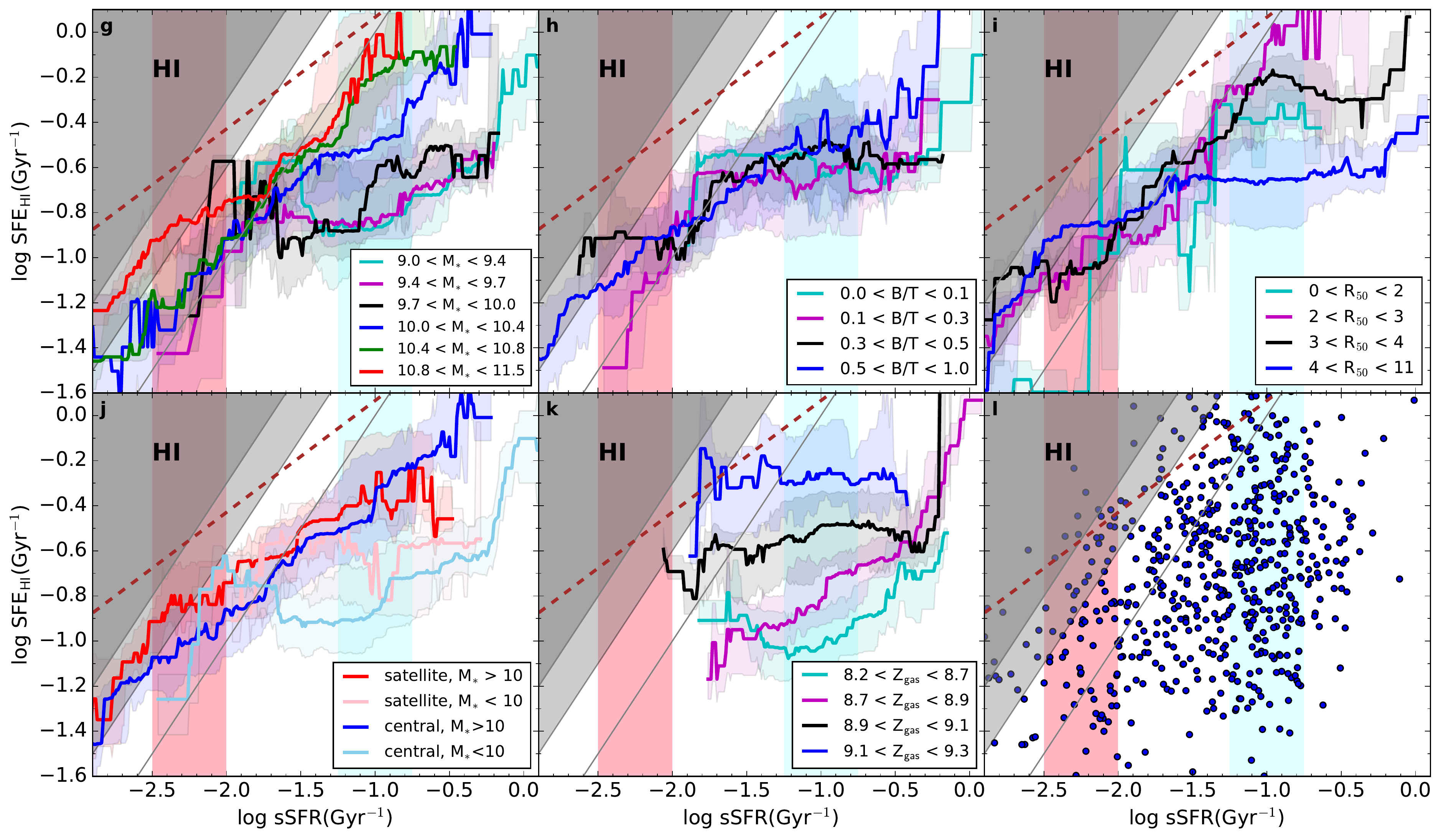}

\caption{The average star formation efficiency as a function of sSFR for molecular gas (upper six panels) and atomic gas (lower six panels). The galaxies are divided into different stellar mass (panel a and g), bulge-to-total ratio (panel b and h), effective radius (panel c and i), centrals and satellites (panel d and j) and different gas-phase metallicities (panel e and k). In panel f and l, we also show the distribution of all galaxies in the \htwo~sample (330 in total) and \hi~sample (662 in total), respectively, without splitting into subsamples. The two dark diagonal lines in the upper six panels mark the constant \htwo~gas to stellar mass ratio of, from left to the right, 1.5\% and 2.5\%, respectively (see text for details). The three dark diagonal lines in the lower six panels mark the constant \hi~gas to stellar mass ratio of, from left to the right, 2\%, 4\% and 10\%, respectively (see text for details). In each panel, the light blue and red shades indicate the position of the star-forming main sequence and quiescent galaxies with $M_*=10^{10}\Msol$, respectively. The brown dashed line in each panel indicates the best ODR fitted relation between SFE and sSFR for \htwo, given by equation (1). Error bars on each line indicate the 32\% and 64\% percentile of the distribution.}
 \label{h2hi}
 \vspace{1pt}
\end{figure*}

Various studies have shown that $\tau_{\rm dep,H_2}$ is tightly correlated with the SFR offset from the MS (\citealp[e.g.,][]{Saintonge:2011ey,2012Saintonge,2017Saintonge,Huang:2014ko,Genzel:2015fq,2017ApJ...837..150S}; \citealp[for a review]{Tacconi:2018jb,2020Tacconi}; \citealp{2020MNRAS.491...69E}). In \citetalias{2021Dou}, we show the $\tau_{\rm dep,H_2}$-sSFR relation is part of the FFR. Similar to Figure 4 in \citetalias{2021Dou}, Figure \ref{h2hi} shows the average SFE as a function of sSFR for \htwo\ (upper six panels) and for \hi\ (lower six panels). We split the galaxies into different stellar mass (panel a and g), bulge-to-total ratio (panel b and h) and effective radius (panel c and i), centrals and satellites (panel d and j), and different gas-phase metallicities (panel e and k). In panel f and l, we also show the distribution of all galaxies in the \htwo\ and \hi\ sample, respectively, without splitting into subsamples. The vertical light blue and red shades in each panel indicate the position of the star-forming MS and the approximate position of the quenched galaxies at $M_*=10^{10}\Msol$, respectively.

Since SFE = sSFR/$\mu$, where $\mu$ is the gas to stellar mass ratio, for a given position on the SFE-sSFR plane, the value of $\mu$ (or the gas fraction $f_{\rm gas}$) is uniquely determined. At a given sSFR, a larger SFE corresponds a lower gas fraction. The two dark diagonal lines in the upper six panels mark the constant $\mu_{\rm H_2}$ of 1.5\% and 2.5\%, respectively. These are the detection limits for the xCOLD GASS galaxies at different stellar mass (1.5\% for log $M_*>$ 10 and 2.5\% for $9 <$ log $M_*<10$). The three dark diagonal lines in the lower six panels mark the constant $\mu_{\rm HI}$ of 2\%, 4\% and 10\%, respectively. These are the detection limits for the xGASS galaxies at different stellar mass (see Section 2). The dark shades in each panel mark the regions below the detection limit for different $M_*$.

The sample selection effects in xCOLD GASS and the uncertainties in SFR measurements have been discussed in detail in \citetalias{2021Dou}. The average \htwo\ detection ratio is higher than 80\% at log sSFR $> -1.6$ Gyr$^{-1}$, where the results should be reliable. For \hi, as in Figure A1 and A2, the average \hi\ detection ratio in xGASS is higher than 80\% at log sSFR $> -1.5$ Gyr$^{-1}$ (where the results should be reliable). For galaxies with log sSFR $< -1.5$ Gyr$^{-1}$, the detection ratio for both \htwo\ and \hi\ decreases rapidly due to the observation limit of the surveys. Hence results in this regime are for \htwo\ or \hi\ detected galaxies, which may not represent the statistics of a unbiased sample. We also note that, as shown in the upper six panels in Figure \ref{h2hi}, the general trend of the SFE$_{\rm H_2}$-sSFR relation seems to persist below log sSFR $< -1.5$ Gyr$^{-1}$, i.e. keep a similar slope and is still independent of other parameters. In the modeling part below, we assume this is true. Future deeper surveys will provide critical observation evidences in the low sSFR regime. If the SFE$_{\rm H_2}$-sSFR relation does change its slope and normalization in the low sSFR regime, or becomes dependent on other galaxies properties, the derived gas cycle during quenching should be revised accordingly.

The results shown in Figure \ref{h2hi} confirm the results in \citetalias{2021Dou} that there exists a tight correlation between SFE$_{\rm H_2}$ and sSFR. This relation holds from star-bursting galaxies to those in the process of being quenched, probing most accurately galaxies with log sSFR $\gtrsim -2$ Gyr$^{-1}$. It is independent of all key parameters that we have explored, including stellar mass, B/T, size, environment (in terms of central/satellite) and gas-phase metallicity. However, no such a tight correlation is found for \hi\ gas. The brown dashed line in all panels indicates the best orthogonal distance regression (ODR) fitted relation between SFE$_{\rm H_2}$ and sSFR as in \citetalias{2021Dou}:
\begin{equation}
\label{eq1}
\rm log\ SFE_{H_2} = 0.50\ log\ sSFR\ + 0.56,
\end{equation}
the 1 $\sigma$ systematic uncertainties in the two fitting parameters are 0.5$\pm$0.021 and 0.56$\pm$0.027, respectively. The slope of 0.5 is in good agreement with previous studies \citep[e.g.,][]{2012Saintonge,Sargent:2014ky,Genzel:2015fq,Tacconi:2018jb}. For the question of whether changes in star formation are primarily driven by the change in SFE or gas fraction, it becomes immediately clear with the equation sSFR = $\mu \times$SFE. The answer is given by the logarithmic slope of the $\mu_{\rm H_2}$-sSFR relation, or equivalently, of the SFE$_{\rm H_2}$-sSFR relation.
Given sSFR = $\mu_{\rm H_2} \times$ SFE$_{\rm H_2}$, the sum of the two logarithmic slopes should be unity. As shown in \citetalias{2021Dou} and equation (1), the measured logarithmic slope is $\sim$ 0.5 for SFE$_{\rm H_2}$-sSFR, $\sim$ 0.6 for $\mu_{\rm H_2}$-sSFR. Hence their sum is indeed about unity. The two logarithmic slopes are similar ($\sim$ 0.5), which means the change in log sSFR is due to similar contribution from the change in log SFE$_{\rm H_2}$ and log $\mu_{\rm H_2}$, i.e. $\mu_{\rm H_2}$ and SFE$_{\rm H_2}$ are comparably important in determining the star formation level (in terms of sSFR), for both quenching and star formation enhancement. Since the exact values of the two slopes may depend on sample selection, observation limit, CO conversion factor, and also fitting method, it remains to be determined more precisely with further surveys.

It should be noted that knowing SFR alone is not enough to know the star formation status of the galaxy, i.e. it is star-forming or quenching. The SFR of a massive galaxy in quenching may be higher than the SFR of a star-forming low-mass galaxy. In addition to the SFR, one also needs to know the stellar mass (hence the sSFR). The same argument also applies to $M_{\rm H_2}$, i.e. a higher $M_{\rm H_2}$ alone does not necessarily mean the galaxy is a star-forming MS galaxy, while a higher $M_{\rm H_2}$/$M_{*}$ (i.e. $\mu$) does, at least in a statistical sense via the observed average $\mu$-sSFR relation. Therefore, the specific relation sSFR = $\mu_{\rm H_2} \times$ SFE$_{\rm H_2}$ is more effective in studying the star formation status and its determining factor.

Interestingly, the sSFR depending on both SFE and $\mu$ holds also locally as derived from the resolved data, as illustrated in Section 4.2 in \citet{2020MNRAS.496.4606M} and the various studies of ALMaQUEST \citep{2020MNRAS.492.6027E,2020MNRAS.493L..39E}. Besides, in the lower panels, all the curves are essentially below the dashed line, which indicates, as mentioned before, on average the SFE$_{\rm H_2}$ is higher than SFE$_{\rm HI}$.

The difference between the SFE-sSFR relation for \hi\ and that for \htwo\ is dramatic. It is well known that the extended \hi\ gas is very sensitive to environmental effects for satellites \citep{1985ApJ...292..404G,2013MNRAS.436...34C}, while for massive central disk galaxies, it remains largely unchanged during quenching \citep{2019ApJ...884L..52Z}. Therefore, different processes, in particular quenching processes (e.g., the two separated quenching channels as in \citealp{Peng:2010gn}, mass quenching and environment quenching) may have very different effects on the \hi\ gas and hence produce very different SFE$_{\rm HI}$-sSFR relations. On the contrary, the SFE$_{\rm H_2}$-sSFR relation remains the same for galaxies with vastly different stellar masses, structures, sizes and in different environments. It should be noted that, this does not mean different processes (e.g., quenching, keeping galaxies evolving on the MS or star-burst triggering) have zero effect on the \htwo\ gas. As we will show in the next subsection, different processes require different \htwo\ gas supply or removal rate, which controls how the galaxy evolves on the (same) SFE$_{\rm H_2}$-sSFR relation (i.e. positions on the relation, evolution speed, evolving to the low or high sSFR direction).

\subsection{Galaxy evolution on the SFE$_{H_2}$-sSFR(t) relation} \label{sec:model}

As discussed in \citetalias{2021Dou}, the scatter of the SFE$_{\rm H_2}$-sSFR relation is 0.2 dex, similar to the combined measurement errors of SFE and sSFR on the orthogonal direction to the fitted line which is 0.21 dex, i.e. the scatter of the SFE$_{\rm H_2}$-sSFR relation can be entirely explained by the measurement errors. This means the intrinsic scatter of this relation is extremely small. This is supported by the fact that the SFE$_{\rm H_2}$-sSFR relation is independent of all key galaxy properties that we have explored in the previous section. This unique feature of the relation makes it simpler and more convenient in modeling the evolution of different galaxy populations than other scaling relations. This is because when galaxy evolving with time, its stellar mass increases due to star formation. If a mass-dependent scaling relation is used, for instance the integrated SFR-$M_{\rm H_2}$ relation (as shown in Figure 3 in \citetalias{2021Dou}), then different SFR-$M_{\rm H_2}$ relations have to be used at different times (due to its strong dependence on stellar mass). Conversely, the SFE$_{\rm H_2}$-sSFR relation is stellar mass independent, hence there is no need to consider this effect. The same argument is also applied to other galaxy properties, such as structure, size and environment that are all changing with time. For instance, as shown in \citetalias{2021Dou} and Figure \ref{h2hi}, the SFR-$M_{\rm H_2}$ relation also depends on B/T and the B/T for a given galaxy is also changing with time. On the contrary, the SFE$_{\rm H_2}$-sSFR relation is independent of all other galaxy properties that we have explored, and therefore we consider it more \textit{fundamental} that the SFR-$M_{\rm H_2}$ relation.

Since the sSFR is directly related to the cosmic time $t$ and acts as the cosmic clock \citep{Peng:2010gn,Peng:2014hn}, the SFE$_{\rm H_2}$-sSFR relation is in fact a SFE$_{\rm H_2}$-sSFR $\sim$ $t$ relation. For simplicity in modeling, we further assume that galaxies with different properties (e.g., stellar masses, metallicities, structures, sizes and in different environments) are all strictly evolving on the SFE$_{\rm H_2}$-sSFR relation given by equation (1). Then we use the gas regulator model to study how different galaxy populations evolve on this scaling relation.

The change of the total cold gas (i.e. \hi\ + \htwo) mass of the galaxy per unit time is given by equation (8) in \citet{Peng:2014hn}. In a similar way, given the mass conservation of \htwo, the change of the \htwo\ gas mass is given by the net \htwo\ supply (or removal) rate minus the \htwo\ consumed by star formation, i.e.,

\begin{align}
\frac{{\rm d}{M_{\rm H_2}}}{{\rm d}{t}} & = \Phi_{\rm H_2} - \Psi_{\rm H_2} - (1-R) \times \rm SFR \nonumber\\
& =\lambda_{\rm net,H_2} \times {\rm SFR} - (1 - R) \times \rm SFR \nonumber\\
& =(\lambda_{\rm net,H_2} - (1- R)) \times {\rm SFE_{H_2}} \times {M_{\rm H_2}}
\label{eq2}
\end{align}
where $R$ is the fraction of the mass of the newly formed stars as measured by the SFR, which is quickly returned to the interstellar medium and is assumed to be about 0.4. Here we have also assumed that stars form only from \htwo\ gas, i.e. (1 - $R$) $\times$ SFR is the total mass permanently consumed by star formation and is also the \htwo\ mass permanently locked up into long-lived stars. $\Phi_{\rm H_2}$ is \htwo\ supply rate and $\Psi_{\rm H_2}$ is \htwo\ removal rate. Hence $\lambda_{\rm net,H_2} = (\Phi_{\rm H_2} - \Psi_{\rm H_2}$) / SFR is the net \htwo\ supply loading factor. It should be noted that we use the term “\htwo\ supply/removal” instead of “\htwo\ inflow/outflow”. This is because \htwo\ does not typically accrete onto galaxies or haloes as “inflow”, except for mergers, but is rather formed inside galaxies from pre-accreted \hi. Hence the “\htwo\ supply” here includes \htwo\ formed inside galaxies, and those brought in by mergers or accreting filaments. The “\htwo\ removal” here includes the \htwo\ outflow driven by star formation or AGN feedback, and it can also include \htwo\ destroyed by photo-dissociation \citep[see e.g.,][]{2006Blitz,2010McKee,2021Morselli}, or removed by external environment effects (e.g., strong ram-pressure stripping).

As noted above, if galaxies with different properties are all assumed to evolve strictly on the SFE$_{\rm H_2}$-sSFR relation given by equation (1), given the analytic form of equation (2), the evolution locus of a galaxy on the SFE$_{\rm H_2}$-sSFR($t$) relation is uniquely controlled by $\lambda_{\rm net,H_2}$. A positive value of $\lambda_{\rm net,H_2}$ means net \htwo\ supply, and a negative value means net \htwo\ removal. In particular, when $\lambda_{\rm net,eq}$ = $(1 - R) \sim$ 0.6, d$M_{\rm H_2}$/d$t$ = 0 and the model achieves equilibrium at a constant \htwo\ gas mass. Therefore, we can use the evolution locus of a given galaxy population on the SFE$_{\rm H_2}$-sSFR($t$) relation to accurately determine $\lambda_{\rm net,H_2}$ and the net \htwo\ gas supply or removal rate.

\begin{figure*}[htbp]
    \begin{center}
       \includegraphics[width=180mm]{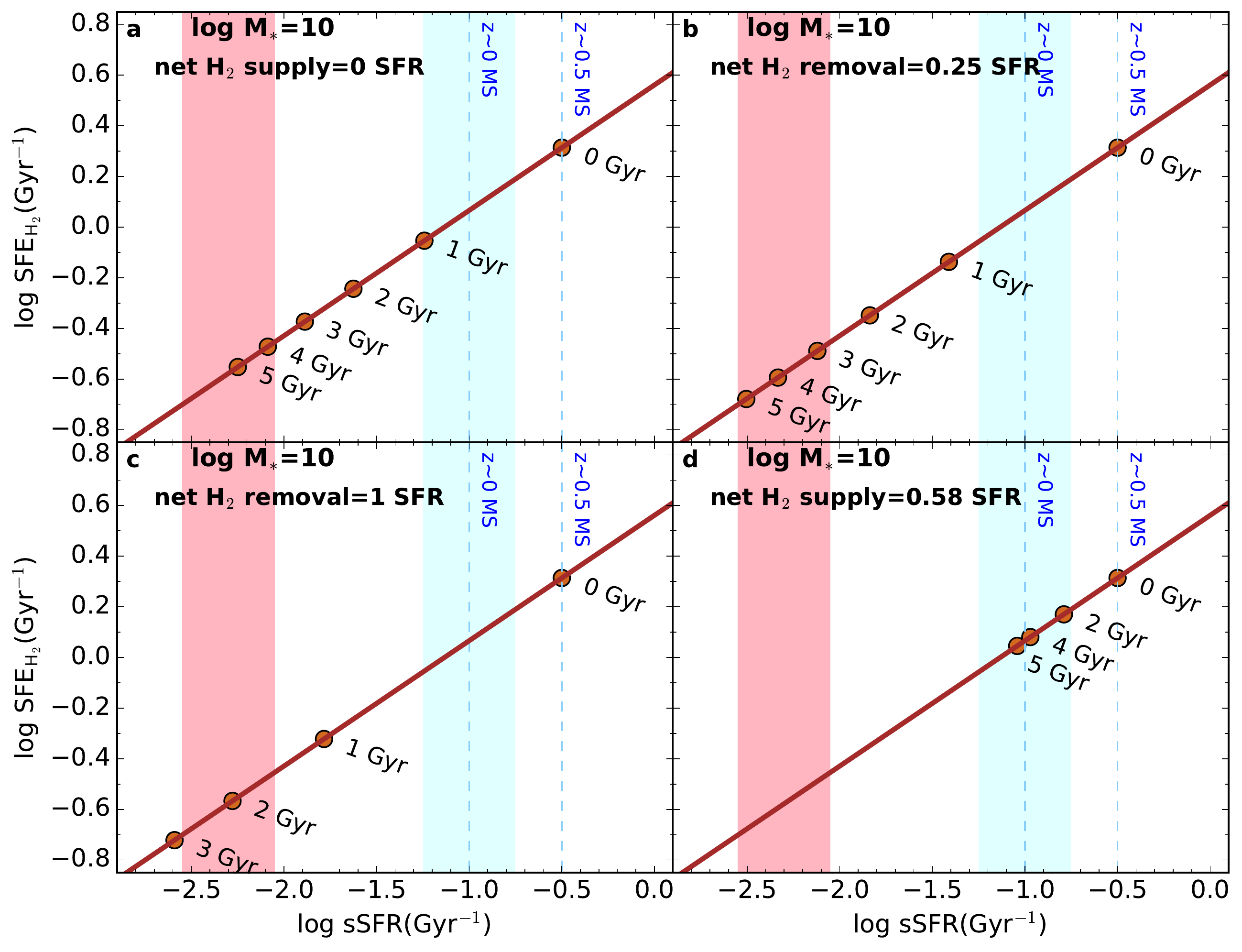}
    \end{center}
\caption{The evolution locus of a model galaxy on the SFE$_{\rm H_2}$-sSFR($t$) relation (thick brown line, given by equation (1)), with an initial $M_*=10^{10}\Msol$ and different values of $\lambda_{\rm net,H_2}$. In each panel, the two vertical blue dashed lines indicate the sSFR of MS at $M_*=10^{10}\Msol$, at $z = 0$ and 0.5. The blue and red shades indicate the approximate position of the star-forming galaxies and fully quenched galaxies at $z \sim 0$, respectively.}
 \label{model_10}
\end{figure*}

\begin{figure*}[htbp]
    \begin{center}
       \includegraphics[width=180mm]{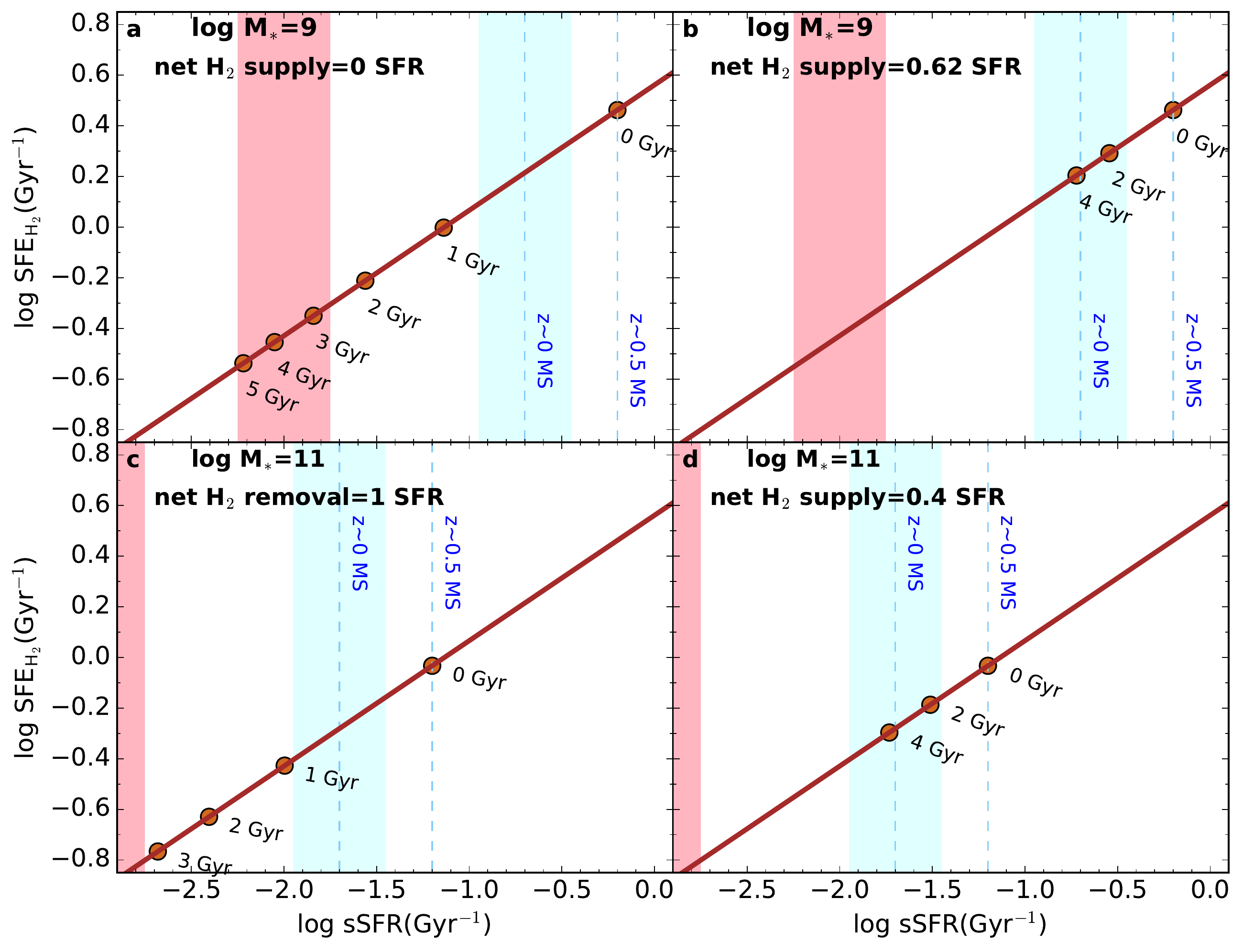}
    \end{center}
\caption{As for Figure \ref{model_10}, but for different initial stellar mass $M_*=10^{9}\Msol$ (upper panels) and $M_*=10^{11}\Msol$ (lower panels), with different values of $\lambda_{\rm net,H_2}$. The left two panels show the cases of quenching, and the right two panels show the cases of evolving on the main sequence.}
 \label{model_9_11}
\end{figure*}

Figure \ref{model_10} and \ref{model_9_11} show the evolution locus of model galaxies with different stellar mass and $\lambda_{\rm net,H_2}$ (labeled on the top of each panel) on the SFE$_{\rm H_2}$-sSFR($t$) relation (thick brown line). Although the SFE$_{\rm H_2}$-sSFR($t$) relation itself does not depend on stellar mass, the sSFR of the MS do. For the MS as shown in Figure \ref{h2hi_2d}, the sSFR at log $M_* \sim$ 9, 10 and 11 is -0.7, -1 and -1.7 Gyr$^{-1}$, respectively. These are indicated by the vertical dashed line, labeled as “$z \sim 0$ MS” in each panel of Figure \ref{model_10} (for log $M_* \sim$ 10) and Figure \ref{model_9_11} (for log $M_* \sim$ 9 and 11). As shown in Figure 3b in \citet{Peng:2015bq}, the stellar age difference between the star-forming and quiescent galaxies is almost constant across the entire observable stellar mass range in SDSS and is equal to $\Delta$age $\sim$ 4 Gyr. This is in broad agreement with other quenching timescales measured in \citet{2016MNRAS.456.4364B,2017ApJ...835..153F,2017ApJ...841L..22G}. Therefore, in the toy model we start the evolution from the MS at $z =$ 0.5 (that is, 4 Gyr ago), shown by the circle dot labeled as “0 Gyr” in each panel. The observed sSFR of the MS at $z =$ 0.5 is about 0.5 dex higher than the MS at $z =$ 0, derived from sSFR $\sim (1+z)^3$ \citep[e.g.,][]{Lilly:2013ko,Speagle:2014dd,2015A&A...579A...2I}.

The initial conditions are the initial $M_*$, redshift and the assumed $\lambda_{\rm net,H_2}$ (labeled in each panel in Figure \ref{model_10}). As mentioned above, the appearance of $M_*$ in the initial conditions is due to the sSFR of the MS is mass-dependent at a given redshift. The SFE$_{\rm H_2}$-sSFR relation itself is mass-independent. One great feature of this approach is that there is no need to assume any initial $M_{\rm H_2}$ or SFR. The evolution of the model galaxy is calculated as follows. The sSFR of the MS is given by the observed sSFR$_{\rm MS}$($M_*$, $z$). Then SFE$_{\rm H_2}$ is given by the SFE$_{\rm H_2}$-sSFR relation in equation (1). $\mu_{\rm H_2}$ is then given by sSFR/SFE$_{\rm H_2}$. Then $\mu_{\rm H_2}$ and $M_*$ together give $M_{\rm H_2}$. Putting SFE$_{\rm H_2}$, $M_{\rm H_2}$ and $\lambda_{\rm net,H_2}$ into equation (2) gives the change of $M_{\rm H_2}$ during d$t$. Repeating this iteration gives the evolution loci of the model galaxy under different assumed gas cycle regime (i.e. different $\lambda_{\rm net,H_2}$) in each panel in Figure \ref{model_10}.

As shown in Figure 1 in \citet{Peng:2014hn}, the average accretion rate of a given dark matter halo and the implied baryonic accretion rate change little in the past few Gyrs (e.g., from $z =$ 0.5 to 0). The \htwo\ supply may also have not changed much, hence a constant $\lambda_{\rm net,H_2}$ has been assumed here. At $z >$ 1, both halo and baryonic accretion rate change rapidly with time, hence a time-dependent $\lambda_{\rm net,H_2}$ will need to be used.

As shown in Figure 3a and 3b, with $\lambda_{\rm net,H_2}$ = 0 (no net \htwo\ supply or removal, like a closed box) or -0.25 (weak net \htwo\ removal), the galaxy evolves rapidly along the SFE$_{\rm H_2}$-sSFR($t$) relation towards low SFE$_{\rm H_2}$ and low sSFR. It will be fully quenched and join the quiescent galaxy population (whose SFRs are to be 1 dex below that of the MS galaxies) within 4-5 Gyr, in good agreement with the observed $\Delta$age $\sim$ 4 Gyr. This suggests that a strong \htwo\ removal is not required to quench the low-redshift MS galaxies. As shown in panel c, with a stronger \htwo\ removal $\lambda_{\rm net,H_2}$ = 1, the galaxy will be fully quenched in about 2 Gyr. However, as shown in Figure 4 of \citet{Peng:2015bq} \citep[see also in][]{2020MNRAS.491.5406T}, with a net gas removal of $\lambda_{\rm net,H_2}$ = 1, the stellar metallicity enhancement during quenching will be significantly smaller than the observed values, and hence argues against strong net outflow during quenching for middle to low-mass galaxies, on average.

As shown in Figure 3d, with $\lambda_{\rm net,H_2}$ = 0.58 (modest net \htwo\ supply), the galaxy evolves slowly downwards along the SFE$_{\rm H_2}$-sSFR($t$) relation. After about 4-5 Gyr (the cosmic time difference between $z =$ 0.5 and 0), it arrives at the MS at $z = 0$. This suggests that to keep the galaxy evolving on the MS, it needs a net \htwo\ gas supply rate of about 0.58 SFR (i.e. $\lambda_{\rm net,H_2}$ = 0.58), which interestingly is about the equilibrium value predicted in the gas regulator model, of $\lambda_{\rm net,eq} = (1-R) \sim$ 0.6 with R assumed to be 0.4. Therefore, the MS galaxies with log $M_*$ = 10 are indeed evolving around the equilibrium state, in which their \htwo\ gas mass is about constant with time. Since its stellar mass will continue to grow due to star formation, their sSFR and gas fraction will still decrease gradually with time.

We investigate the effect of the uncertainties in the fitting parameters in equation (1) on the results, by allowing the two fitting parameters to change within the estimated uncertainties. We also test our results using fitting parameters derived from ordinary least squares, which produces a slightly shallower slope of 0.44. These results are shown in Figure A3 in the Appendix. The evolution loci of the model galaxy on different lines are all similar. Hence the uncertainties in the fitting parameters only have small effect on our results.

Figure \ref{model_9_11} is similar to Figure \ref{model_10}, but for galaxies with log $M_*$ = 9 (upper panels) and log $M_*$ = 11 (lower panels). As mentioned above, the difference due to a different $M_*$ is the sSFR of the MS, which is now -0.7 Gyr$^{-1}$ for log $M_* \sim$ 9 and -1.7 Gyr$^{-1}$ for log $M_* \sim$ 11 at $z \sim$ 0. The case for log $M_* \sim$ 9 is very similar to log $M_* \sim$ 10 as shown in Figure \ref{model_10}. With no net \htwo\ supply, the galaxy will be fully quenched in about 4 Gyr. To keep it evolving on the MS, a net \htwo\ supply of $\lambda_{\rm net,H_2} \sim$ 0.6 is required, which is also equal to the equilibrium value of $\lambda_{\rm net,eq} = (1-R)$, see equation (2), with R assumed to be 0.4. For a massive MS galaxies with log $M_* \sim$ 11, the situation is different. To fully quench it in 4 Gyr (i.e. within the observed $\Delta$age $\sim$ 4 Gyr as discussed before), the halt of \htwo\ supply (through either prior strangulation of predominantly \hi\ gas, and/or prevention of \hi-to-\htwo\ conversion inside galaxies) is not enough, and it requires an additional net \htwo\ removal with $\lambda_{\rm net,H_2} = -1$ (i.e. a mass-loading factor of unity). This means the massive galaxies are more resistant to quenching due to the halt of \htwo\ supply. This is because massive MS galaxies have a lower sSFR than low-mass MS galaxies, hence a lower SFE or longer $\tau_{\rm dep}$. While the definition of quench is the same for all galaxies, i.e. to decrease the sSFR by more than 1 dex relative to the MS. This effect is clearly seen in other panels that the decrease of sSFR is largest during the first Gyr of quenching. For the same reason, it is also relatively easier to keep the massive galaxy stay on the MS, which requires a net \htwo\ supply of $\lambda_{\rm net,H_2} \sim$ 0.4, lower than the $\lambda_{\rm net,eq}$. This means the \htwo\ gas mass in the massive MS galaxies will slowly decrease with time.

Putting together, to quench middle to low-mass galaxies, the halt of \htwo\ supply is required. The observed stellar metallicity difference between star-forming and passive galaxies argues against any strong outflow or gas removal in this mass range \citep{Peng:2015bq,2020MNRAS.491.5406T}. Massive galaxies are more resistant to quenching due to the halt of \htwo\ supply. To quench them, additional \htwo\ gas removal is required (with a net mass-loading factor about unity), probably driven by AGN feedback in massive galaxies \citep[e.g.,][for a review]{2014Cicone,2019ApJ...875...21F,2020Veilleux} or by photodissociation of \htwo. Interestingly, this does not contradict to the stellar metallicity requirement as in \citet{Peng:2015bq}. This is because although gas outflow or gas removal will act to reduce the stellar metallicity enhancement during quenching, the $f_{\rm gas}$ of the MS massive galaxy on average is already very low and the stellar metallicity enhancement will be very limited even if the galaxy is quenched with zero gas removal. Therefore, counterintuitively, although the low-mass galaxies have a significantly higher $f_{\rm gas}$ in \hi\ than massive galaxies (evidently in Figure 2g), they are more sensitive to the \htwo\ gas supply and are easier to be quenched by the halt of \htwo\ supply (through either prior strangulation of predominantly \hi\ gas due to environmental effect, and/or prevention of \hi-to-\htwo\ conversion inside galaxies).

It is also interesting to notice that for the cases of quenching (i.e. with no \htwo\ supply or net \htwo\ removal, including Figure 3a,b,c and Figure 4a,c), the quenching speed (dsSFR/d$t$) is faster in the beginning due to the higher SFE (hence shorter $\tau_{\rm dep}$). The sSFR drops more than 1 dex in the first 1 or 2 Gyr (as in Figure \ref{model_10} and Figure \ref{model_9_11}).

While keeping galaxies evolving on the MS, the middle to low-mass galaxies require a net \htwo\ supply of $\lambda_{\rm net,H_2} \sim$ 0.6, which is the equilibrium state predicted by the gas regulator model of $\lambda_{\rm net,eq} = (1-R) \sim$ 0.6 with $R$ assumed to be 0.4. Massive galaxies require a slightly lower net \htwo\ supply of $\lambda_{\rm net,H_2} \sim$ 0.4, which is smaller than the $\lambda_{\rm net,eq}$ due to their relatively lower sSFR.

It should be appreciated that in the simple framework of FFR, the value of $M_*$ only matters due to its (weak) dependence on sSFR (at a given $z$). As a consequence of this, at a given $z$, the MS galaxies with different $M_*$ locate on slightly different positions on the SFE$_{\rm H_2}$-sSFR($t$) relation. Apart from this, their evolution on the SFE$_{\rm H_2}$-sSFR($t$) relation is controlled solely by $\lambda_{\rm net,H_2}$. A zero or negative $\lambda_{\rm net,H_2}$ (net \htwo\ removal) in general quenches the galaxy and makes it rapidly evolve to join the quiescent population within 4-5 Gyr; a positive $\lambda_{\rm net,H_2}$ around $\lambda_{\rm net,eq}$ (modest net \htwo\ supply) can keep the galaxy evolve on the MS (with slowly decreasing sSFR), and a large positive $\lambda_{\rm net,H_2}$ (strong net \htwo\ supply introduced by e.g., mergers or strong dynamical interactions) can boost the sSFR (e.g., trigger starbursts).

We stress that although the SFE$_{\rm H_2}$-sSFR relation is very tight and is independent of other key galaxy properties such as $M_*$, structure, metallicity and environment, the value of $\lambda_{\rm net,H_2}$ may strongly depends on these properties. For instance, when move from fields to groups, and to clusters, as environmental quenching becomes more and more effective, the value of $\lambda_{\rm net,H_2}$ is expected to become progressively smaller (i.e. suppressed \hi\ gas inflow, hence less available \hi\ gas to be converted into \htwo), then around zero (i.e. strangulation of \hi\ gas inflow and no more \htwo\ supply), and then negative (i.e. strangulation plus additional gas removal, such as gas stripping in clusters). This means that galaxies in dense regions, on average, will evolve faster along the SFE$_{\rm H_2}$-sSFR relation to the low sSFR direction (i.e. being quenched faster). Therefore, the fact that SFE$_{\rm H_2}$-sSFR relation is independent of environment does not contradict to the well-known fact that the fraction of the passive galaxies is higher in dense regions \citep{2004MNRAS.353..713K,2006MNRAS.373..469B,Peng:2010gn}. This argument can also be applied to explain morphological quenching \citep{2009ApJ...707..250M,2014ApJ...785...75G,2020MNRAS.495..199G}, i.e. the SFE$_{\rm H_2}$-sSFR relation is independent of B/T, but the galaxies with a more massive bulge (hence higher B/T) may have a smaller $\lambda_{\rm net,H_2}$, and hence more suppressed star formation.   \\[10pt]

\subsection{Galaxy evolution on the SFE$_{HI}$-sSFR relations?} \label{sec:model}

As shown in Figure \ref{h2hi}, the SFE$_{\rm HI}$-sSFR relation shows strong systematic dependence on all key parameters that we have explored (to a less extent for B/T), in particular at the star-forming regions at log sSFR $> -1.5$  (where the sample has a high detection rate $>$ 80\%). We note that at log sSFR $< -1.5$, the \hi\ detection rate drops rapidly below 20\% (Figure A2). This is evident in panel g,h,i,j in Figure \ref{h2hi} that the differences between different curves become smaller at lower sSFR, which are likely caused by the increasing non-detections in \hi. The true differences could be much larger, like those at log sSFR $> -1.5$. Comparing panel f and l, the scatter of the SFE-sSFR relation for \hi\ is significantly larger than that for \htwo, in particular at log sSFR $>$ -1.5, where both the \hi\ and \htwo\ sample have a high detection rate $>$ 80\%. In general, given the fact that the SFE$_{\rm HI}$-sSFR relation is very different for galaxies with different $M_*$, structures, metallicities and in different environments (i.e. central/satellite), it is impossible to derive a simple \hi\ gas supply/removal rate as for \htwo. As discussed before, this reflects that different quenching mechanisms may have very different effects on \hi\ gas and hence produce very different SFE$_{\rm HI}$-sSFR relations. We will further investigate this in the future work.

\section{Summary}

The main focus of this paper is to explore how to use the Fundamental Formation Relation (FFR) of the molecular gas, in combination with the gas regulator model, to study the gas cycle and star formation in the local galaxy populations. Our results and conclusions may be summarized as follows.

(1) At a given stellar mass and SFR, on average, the $M_{\rm HI}$ is higher than the $M_{\rm H_2}$, especially for low-mass galaxies that are \hi\ dominated. Since the typical timescale of \htwo\ formation in equilibrium is much shorter than $\tau_{\rm dep,H_2}$ of typical star-forming MS galaxies. This suggests a significant fraction of \hi\ gas will not be able to form \htwo, in particular in low-mass galaxies.

(2) The \hi\ gas does not follow a similar FFR as the \htwo. The SFE$_{\rm HI}$-sSFR relation shows significant scatter and strong systematic dependence on all of the key galaxy properties that we have explored, while the SFE$_{\rm H_2}$-sSFR relation does not. This dramatic difference between \hi\ and \htwo\ indicates that different processes (e.g., quenching by different mechanisms or star-forming) may have very different effects on the \hi\ gas and hence produce different SFE$_{\rm HI}$-sSFR relations, while SFE$_{\rm H_2}$-sSFR relation remains unaffected. The tighter correlation between sSFR and SFE$_{\rm H_2}$ is also indicative of the well-known fact that stars form predominantly in molecular clouds, not from atomic hydrogen.

(3) The unique feature of the SFE$_{\rm H_2}$-sSFR relation (i.e. very small scatter and independent of other galaxy properties) makes it a simple and convenient tool to study the evolution of different galaxy populations. For simplicity, we further assume that galaxies with different properties are all strictly evolving on the SFE$_{\rm H_2}$-sSFR relation given by equation (1). Then we apply the gas regulator model to study how different galaxy populations evolve on this scaling relation. We show how galaxies evolve on the SFE$_{\rm H_2}$-sSFR relation (e.g., evolution to the low or high sSFR direction, how fast is the evolution) is uniquely controlled by a single parameter $\lambda_{\rm net,H_2}$. To keep galaxies evolving on the star-forming MS, the middle to low-mass galaxies requires a net \htwo\ supply of $\lambda_{\rm net,H_2} \sim$ 0.6, which is the equilibrium state predicted by the gas regulator model of $\lambda_{\rm net,eq} = (1-R)$ with $R$ assumed to be 0.4. Massive galaxies require a lower net supply of $\lambda_{\rm net,H_2} < \lambda_{\rm net,eq}$ to stay on the MS.

(4) To quench middle to low-mass galaxies, the halt of \htwo\ supply is sufficient (through either prior strangulation of predominantly \hi\ gas due to environmental effect, and/or prevention of \hi-to-\htwo\ conversion inside galaxies). The observed stellar metallicity difference argues against any strong net outflow or gas removal in this mass range. To quench massive galaxies, additional \htwo\ gas removal (with a mass-loading factor of about unity, probably driven by AGN feedback) is required, i.e. massive galaxies are more resistant to quenching due to the halt of \htwo\ supply, despite their lower gas fraction than galaxies with lower $M_*$. For all galaxies, quenching is faster in the beginning, and the sSFR drops more than 1 dex in the first 1 or 2 Gyr's quenching.

(5) Although the SFE$_{\rm H_2}$-sSFR relation is very tight and is independent of all key galaxy properties that we have explored, the value of $\lambda_{\rm net,H_2}$ may strongly depends on these properties. For instance, the value of $\lambda_{\rm net,H_2}$ is expected to become progressively smaller, when move from field to dense regions where environmental quenching becomes more effective. This means that galaxies in dense regions, on average, will evolve more rapidly along the SFE$_{\rm H_2}$-sSFR relation to the low sSFR direction (i.e. being quenched faster). This argument can also be applied to explain other phenomenons, such as morphological quenching, i.e. the SFE$_{\rm H_2}$-sSFR relation is independent of B/T, but galaxies with a more massive bulge may have a smaller $\lambda_{\rm net,H_2}$, and hence more suppressed star formation (due to the smaller $\lambda_{\rm net,H_2}$).

We caution that the above results are based on galaxies with \htwo\ and \hi\ detections. Due to the observation limit of the surveys, the detection ratio for both \htwo\ and \hi\ decreases rapidly at log sSFR $< -1.5$ Gyr$^{-1}$. Hence results in this regime may not represent the statistics of a unbiased sample. On the other hand, as shown in the upper six panels in Figure \ref{h2hi}, the general trend of the SFE$_{\rm H_2}$-sSFR relation seems to persist below log sSFR $< -1.5$ Gyr$^{-1}$. Future deeper surveys will provide critical observations in the low sSFR regime.

In the framework of FFR, the facts that the SFE$_{\rm H_2}$-sSFR relation is independent of all key galaxy properties that we have explored, and that sSFR is directly related to the cosmic time and acts as the cosmic clock, make it natural and extremely simple to study how different galaxy populations (with different properties and undergoing different quenching or star-forming processes) evolve on the same SFE$_{\rm H_2}$-sSFR $\sim t$ relation, for instance, to derive their gas supply/removal rate during quenching or staying on MS. Both SFE$_{\rm H_2}$ and sSFR can be observationally determined, and the scatter of the SFE$_{\rm H_2}$-sSFR relation is only 0.2 dex, one of the tightest scaling relations in the field of star formation. Hence the derived \htwo\ supply and removal rates are of high accuracy. As in \citetalias{2021Dou}, we note that both 1/SFE$_{\rm H_2}$ and 1/sSFR are timescales, one for gas and one for stars. 1/SFE is rephrased as “galactic clock” in \citet{2020Tacconi} and 1/sSFR is rephrased as “cosmic clock” in \citet{Peng:2010gn}. Both of them are closely tied to the growth of the dark matter halo. We will further explore their interrelationship and their connections to the halo in our following paper.

\acknowledgments
We gratefully acknowledge the anonymous referee for
comments and criticisms that have improved the paper. We thank Sara L. Ellison, Li-Hwai Lin, and Jing Wang for useful discussions. This work is supported in part by the Natural Science Foundation of China (grants No.11773001, 11721303, 11991052, 11473002, 11690024, 11725313, 11861131007, 11420101002, 11733002 and 11633006), National Key R\&D Program of China (grant No. 2016YFA0400702, 2016YFA0400704, 2017YFA0402704 and 2017YFA0402703), Chinese Academy of Sciences Key Research Program of Frontier Sciences (grant No.QYZDJSSW-SLH008 and QYZDJSSW-SYS008). A.R. acknowledges support from an INAF/PRIN-SKA 2017 (ESKAPE-HI) grant. F.M. acknowledges support from the INAF PRIN-SKA 2017 program 1.05.01.88.04. R.M. acknowledges ERC Advanced Grant 695671 “QUENCH” and support from the Science and Technology Facilities Council (STFC). S.L. acknowledges support from the Swiss National Science Foundation.

\appendix

\begin{figure*}[htbp]
    \begin{center}
       \includegraphics*[scale=0.5]{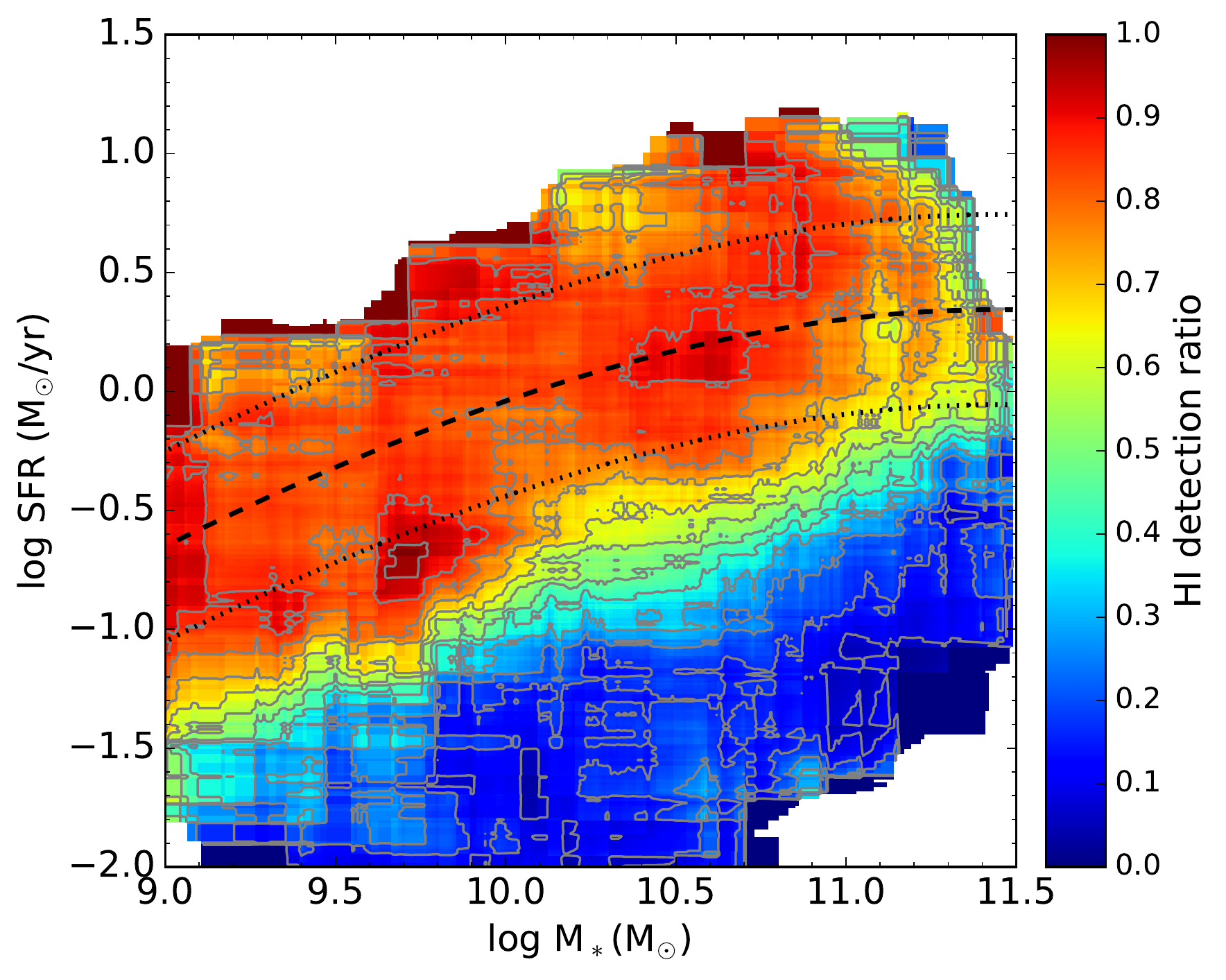}
    \end{center}
\noindent\textbf{Figure A1.} The average \hi\ detection ratio in xGASS sample, as a function of stellar mass ($M_*$) and star formation rate (SFR), determined within moving boxes of size 0.5 dex in mass and 0.5 dex in SFR. Each galaxy is weighted by a correction factor to account for selection effects in stellar mass. The dashed line indicates the position of the star-forming main sequence defined in \citet{2016MNRAS.462.1749S}. The dotted lines indicate $\pm$0.4 dex scatter around the main sequence.
 \label{hi_det_2d}
\end{figure*}

\begin{figure*}[htbp]
    \begin{center}
       \includegraphics*[width=180mm]{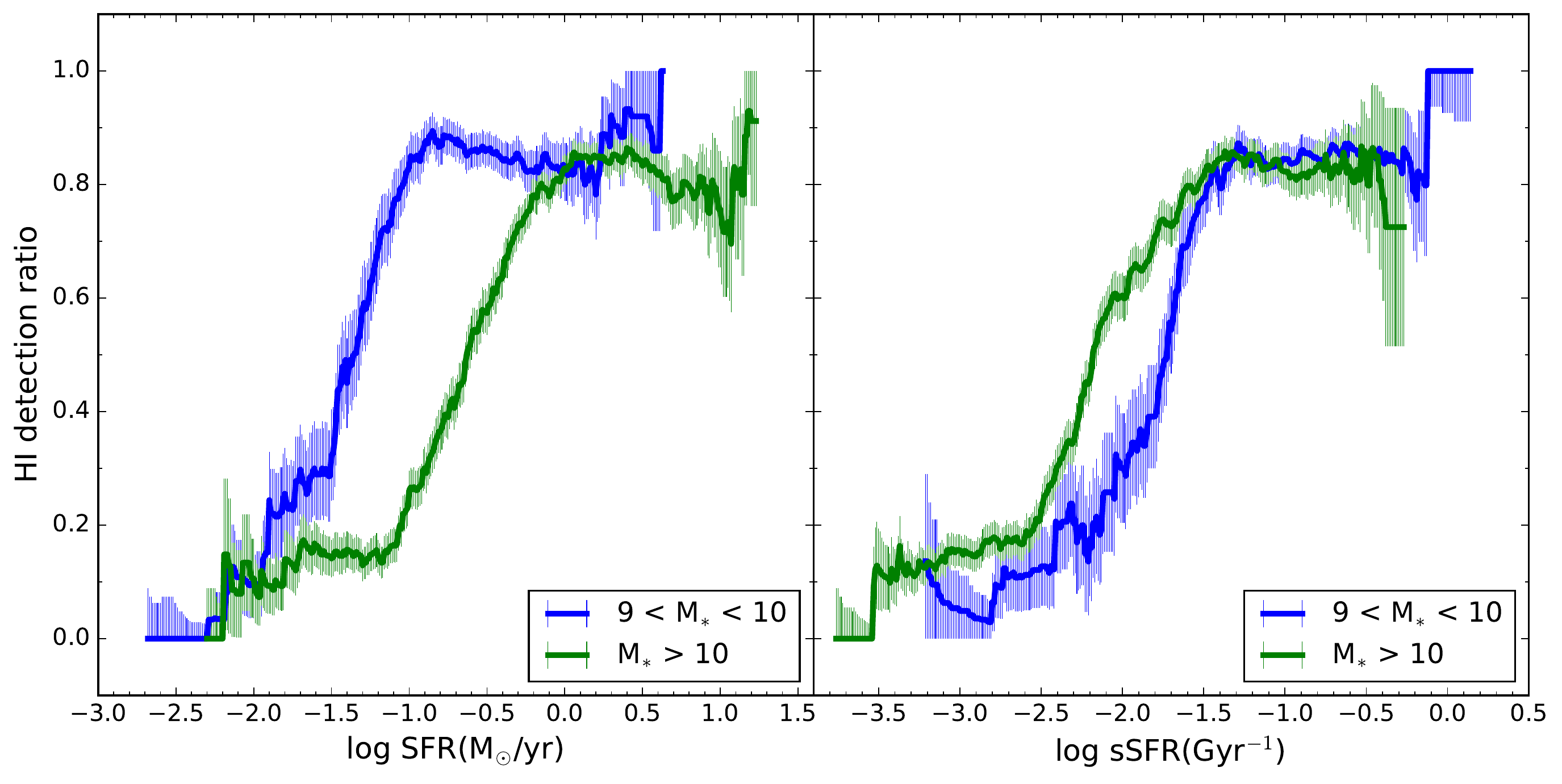}
    \end{center}
\noindent\textbf{Figure A2.} The average \hi\ detection ratio in xGASS sample as a function of SFR (left panel) and sSFR (right panel), for galaxies within different stellar mass bins. The average \hi~detection ratio is larger than 80\% at log sSFR $>-1.5$ Gyr$^{-1}$.
 \label{hi_det_1d}
\end{figure*}

\begin{figure*}[htbp]
    \begin{center}
       \includegraphics*[width=180mm]{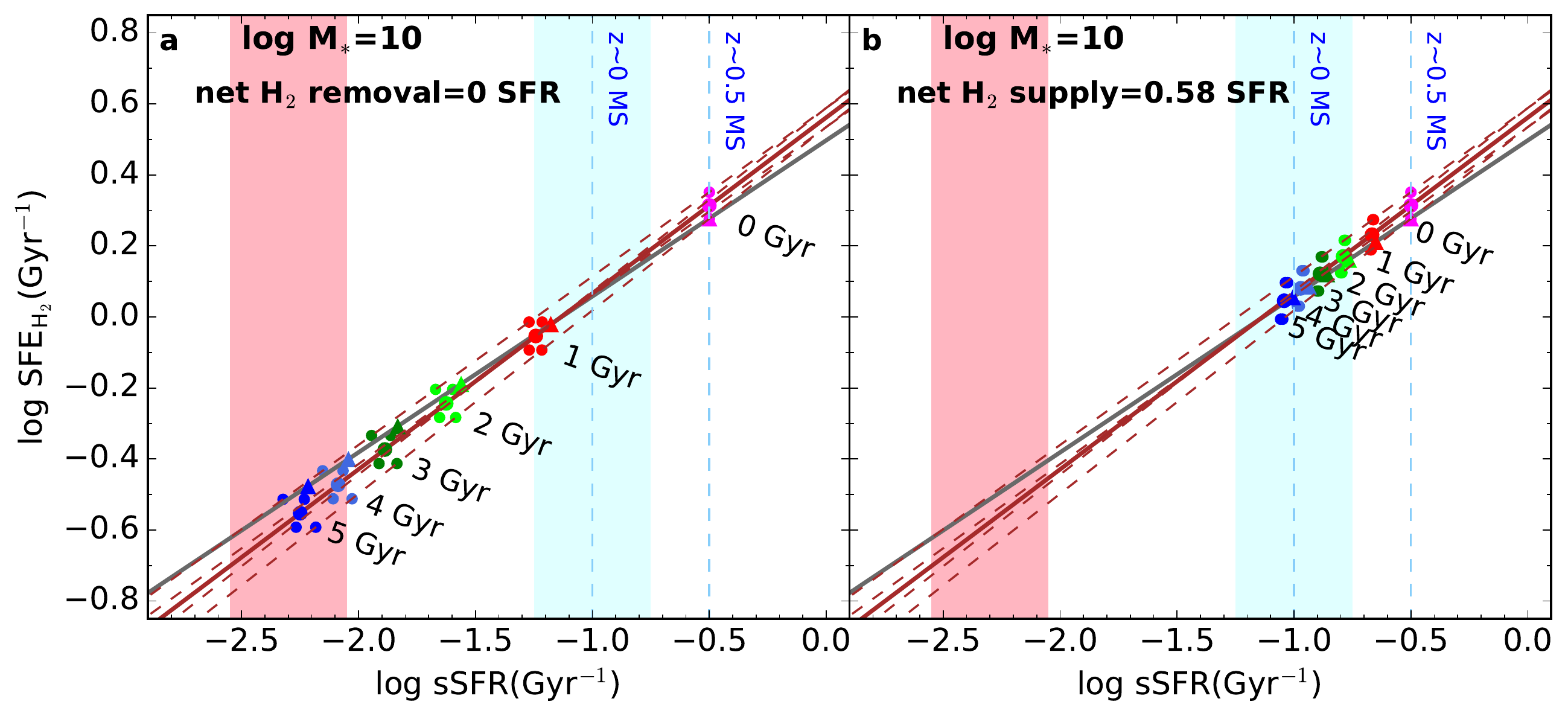}
    \end{center}
\noindent\textbf{Figure A3.} As for Figure \ref{model_10}, but for a model galaxy evolving on different SFE$_{\rm H_2}$-sSFR($t$) relations by taking into account the uncertainties in equation (1). The solid line is given by equation (1). Other four dashed lines are the results of four different combinations of the uncertainties in the two fitting parameters in equation (1). The gray solid line and triangles show the results using ordinary least squares fitting: log SFE$_{\rm H_2}$ = 0.44 ($\pm$0.02) log sSFR + 0.5 ($\pm$0.03). The evolution loci of the model galaxy on different lines are all similar. Evidently, the uncertainties in the fitting parameters only have small effect on the results.
 \label{uncertainties}
\end{figure*}


\clearpage


\begin{thebibliography}{}
\expandafter\ifx\csname natexlab\endcsname\relax\def\natexlab#1{#1}\fi
\providecommand{\url}[1]{\href{#1}{#1}}

\bibitem[{{Abadi} {et~al.}(1999){Abadi}, {Moore}, \&
  {Bower}}]{1999MNRAS.308..947A}
{Abadi}, M.~G., {Moore}, B., \& {Bower}, R.~G. 1999, \mnras, 308, 947

\bibitem[{{Abazajian} {et~al.}(2009){Abazajian}, {Adelman-McCarthy},
  {Ag{\"u}eros}, {Allam}, {Allende Prieto}, {An}, {Anderson}, {Anderson},
  {Annis}, {Bahcall}, {Bailer-Jones}, {Barentine}, {Bassett}, {Becker},
  {Beers}, {Bell}, {Belokurov}, {Berlind}, {Berman}, {Bernardi}, {Bickerton},
  {Bizyaev}, {Blakeslee}, {Blanton}, {Bochanski}, {Boroski}, {Brewington},
  {Brinchmann}, {Brinkmann}, {Brunner}, {Budav{\'a}ri}, {Carey}, {Carliles},
  {Carr}, {Castander}, {Cinabro}, {Connolly}, {Csabai}, {Cunha}, {Czarapata},
  {Davenport}, {de Haas}, {Dilday}, {Doi}, {Eisenstein}, {Evans}, {Evans},
  {Fan}, {Friedman}, {Frieman}, {Fukugita}, {G{\"a}nsicke}, {Gates},
  {Gillespie}, {Gilmore}, {Gonzalez}, {Gonzalez}, {Grebel}, {Gunn},
  {Gy{\"o}ry}, {Hall}, {Harding}, {Harris}, {Harvanek}, {Hawley}, {Hayes},
  {Heckman}, {Hendry}, {Hennessy}, {Hindsley}, {Hoblitt}, {Hogan}, {Hogg},
  {Holtzman}, {Hyde}, {Ichikawa}, {Ichikawa}, {Im}, {Ivezi{\'c}}, {Jester},
  {Jiang}, {Johnson}, {Jorgensen}, {Juri{\'c}}, {Kent}, {Kessler}, {Kleinman},
  {Knapp}, {Konishi}, {Kron}, {Krzesinski}, {Kuropatkin}, {Lampeitl},
  {Lebedeva}, {Lee}, {Lee}, {French Leger}, {L{\'e}pine}, {Li}, {Lima}, {Lin},
  {Long}, {Loomis}, {Loveday}, {Lupton}, {Magnier}, {Malanushenko},
  {Malanushenko}, {Mand elbaum}, {Margon}, {Marriner}, {Mart{\'\i}nez-Delgado},
  {Matsubara}, {McGehee}, {McKay}, {Meiksin}, {Morrison}, {Mullally}, {Munn},
  {Murphy}, {Nash}, {Nebot}, {Neilsen}, {Newberg}, {Newman}, {Nichol},
  {Nicinski}, {Nieto-Santisteban}, {Nitta}, {Okamura}, {Oravetz}, {Ostriker},
  {Owen}, {Padmanabhan}, {Pan}, {Park}, {Pauls}, {Peoples}, {Percival}, {Pier},
  {Pope}, {Pourbaix}, {Price}, {Purger}, {Quinn}, {Raddick}, {Re Fiorentin},
  {Richards}, {Richmond}, {Riess}, {Rix}, {Rockosi}, {Sako}, {Schlegel},
  {Schneider}, {Scholz}, {Schreiber}, {Schwope}, {Seljak}, {Sesar}, {Sheldon},
  {Shimasaku}, {Sibley}, {Simmons}, {Sivarani}, {Allyn Smith}, {Smith},
  {Smol{\v{c}}i{\'c}}, {Snedden}, {Stebbins}, {Steinmetz}, {Stoughton},
  {Strauss}, {SubbaRao}, {Suto}, {Szalay}, {Szapudi}, {Szkody}, {Tanaka},
  {Tegmark}, {Teodoro}, {Thakar}, {Tremonti}, {Tucker}, {Uomoto}, {Vanden
  Berk}, {Vandenberg}, {Vidrih}, {Vogeley}, {Voges}, {Vogt}, {Wadadekar},
  {Watters}, {Weinberg}, {West}, {White}, {Wilhite}, {Wonders}, {Yanny},
  {Yocum}, {York}, {Zehavi}, {Zibetti}, \& {Zucker}}]{Abazajian:2009ef}
{Abazajian}, K.~N., {Adelman-McCarthy}, J.~K., {Ag{\"u}eros}, M.~A., {et~al.}
  2009, \apjs, 182, 543

\bibitem[{{Accurso} {et~al.}(2017){Accurso}, {Saintonge}, {Catinella},
  {Cortese}, {Dav{\'e}}, {Dunsheath}, {Genzel}, {Gracia-Carpio}, {Heckman},
  {Jimmy}, {Kramer}, {Li}, {Lutz}, {Schiminovich}, {Schuster}, {Sternberg},
  {Sturm}, {Tacconi}, {Tran}, \& {Wang}}]{2017Accurso}
{Accurso}, G., {Saintonge}, A., {Catinella}, B., {et~al.} 2017, \mnras, 470,
  4750

\bibitem[{{Baldry} {et~al.}(2006){Baldry}, {Balogh}, {Bower}, {Glazebrook},
  {Nichol}, {Bamford}, \& {Budavari}}]{2006MNRAS.373..469B}
{Baldry}, I.~K., {Balogh}, M.~L., {Bower}, R.~G., {et~al.} 2006, \mnras, 373,
  469

\bibitem[{{Balogh} \& {Morris}(2000)}]{2000MNRAS.318..703B}
{Balogh}, M.~L., \& {Morris}, S.~L. 2000, \mnras, 318, 703

\bibitem[{{Balogh} {et~al.}(2000){Balogh}, {Navarro}, \&
  {Morris}}]{2000ApJ...540..113B}
{Balogh}, M.~L., {Navarro}, J.~F., \& {Morris}, S.~L. 2000, \apj, 540, 113

\bibitem[{{Balogh} {et~al.}(2016){Balogh}, {McGee}, {Mok}, {Muzzin}, {van der
  Burg}, {Bower}, {Finoguenov}, {Hoekstra}, {Lidman}, {Mulchaey}, {Noble},
  {Parker}, {Tanaka}, {Wilman}, {Webb}, {Wilson}, \&
  {Yee}}]{2016MNRAS.456.4364B}
{Balogh}, M.~L., {McGee}, S.~L., {Mok}, A., {et~al.} 2016, \mnras, 456, 4364

\bibitem[{{Belfiore} {et~al.}(2019){Belfiore}, {Vincenzo}, {Maiolino}, \&
  {Matteucci}}]{2019MNRAS.487..456B}
{Belfiore}, F., {Vincenzo}, F., {Maiolino}, R., \& {Matteucci}, F. 2019,
  \mnras, 487, 456

\bibitem[{{Blitz} \& {Rosolowsky}(2006)}]{2006Blitz}
{Blitz}, L., \& {Rosolowsky}, E. 2006, \apj, 650, 933

\bibitem[{{Bouch{\'e}} {et~al.}(2010){Bouch{\'e}}, {Dekel}, {Genzel}, {Genel},
  {Cresci}, {F{\"o}rster Schreiber}, {Shapiro}, {Davies}, \&
  {Tacconi}}]{2010ApJ...718.1001B}
{Bouch{\'e}}, N., {Dekel}, A., {Genzel}, R., {et~al.} 2010, \apj, 718, 1001

\bibitem[{{Catinella} {et~al.}(2010){Catinella}, {Schiminovich}, {Kauffmann},
  {Fabello}, {Wang}, {Hummels}, {Lemonias}, {Moran}, {Wu}, {Giovanelli},
  {Haynes}, {Heckman}, {Basu-Zych}, {Blanton}, {Brinchmann}, {Budav{\'a}ri},
  {Gon{\c{c}}alves}, {Johnson}, {Kennicutt}, {Madore}, {Martin}, {Rich},
  {Tacconi}, {Thilker}, {Wild}, \& {Wyder}}]{2010MNRAS.403..683C}
{Catinella}, B., {Schiminovich}, D., {Kauffmann}, G., {et~al.} 2010, \mnras,
  403, 683

\bibitem[{{Catinella} {et~al.}(2013){Catinella}, {Schiminovich}, {Cortese},
  {Fabello}, {Hummels}, {Moran}, {Lemonias}, {Cooper}, {Wu}, {Heckman}, \&
  {Wang}}]{2013MNRAS.436...34C}
{Catinella}, B., {Schiminovich}, D., {Cortese}, L., {et~al.} 2013, \mnras, 436,
  34

\bibitem[{{Catinella} {et~al.}(2018){Catinella}, {Saintonge}, {Janowiecki},
  {Cortese}, {Dav{\'e}}, {Lemonias}, {Cooper}, {Schiminovich}, {Hummels},
  {Fabello}, {Ger{\'e}b}, {Kilborn}, \& {Wang}}]{2018Catinella}
{Catinella}, B., {Saintonge}, A., {Janowiecki}, S., {et~al.} 2018, \mnras, 476,
  875

\bibitem[{{Chabrier}(2003)}]{2003PASP..115..763C}
{Chabrier}, G. 2003, \pasp, 115, 763

\bibitem[{{Cicone} {et~al.}(2014){Cicone}, {Maiolino}, {Sturm},
  {Graci{\'a}-Carpio}, {Feruglio}, {Neri}, {Aalto}, {Davies}, {Fiore},
  {Fischer}, {Garc{\'\i}a-Burillo}, {Gonz{\'a}lez-Alfonso}, {Hailey-Dunsheath},
  {Piconcelli}, \& {Veilleux}}]{2014Cicone}
{Cicone}, C., {Maiolino}, R., {Sturm}, E., {et~al.} 2014, \aap, 562, A21

\bibitem[{{Conselice} {et~al.}(2013){Conselice}, {Mortlock}, {Bluck},
  {Gr{\"u}tzbauch}, \& {Duncan}}]{2013MNRAS.430.1051C}
{Conselice}, C.~J., {Mortlock}, A., {Bluck}, A. F.~L., {Gr{\"u}tzbauch}, R., \&
  {Duncan}, K. 2013, \mnras, 430, 1051

\bibitem[{{Croton} {et~al.}(2006){Croton}, {Springel}, {White}, {De Lucia},
  {Frenk}, {Gao}, {Jenkins}, {Kauffmann}, {Navarro}, \&
  {Yoshida}}]{2006MNRAS.365...11C}
{Croton}, D.~J., {Springel}, V., {White}, S. D.~M., {et~al.} 2006, \mnras, 365,
  11

\bibitem[{{Dav{\'e}} {et~al.}(2012){Dav{\'e}}, {Finlator}, \&
  {Oppenheimer}}]{2012MNRAS.421...98D}
{Dav{\'e}}, R., {Finlator}, K., \& {Oppenheimer}, B.~D. 2012, \mnras, 421, 98

\bibitem[{{Dekel} \& {Birnboim}(2006)}]{2006MNRAS.368....2D}
{Dekel}, A., \& {Birnboim}, Y. 2006, \mnras, 368, 2

\bibitem[{{Dekel} \& {Mandelker}(2014)}]{2014MNRAS.444.2071D}
{Dekel}, A., \& {Mandelker}, N. 2014, \mnras, 444, 2071

\bibitem[{{Dou} {et~al.}(2021){Dou}, {Peng}, {Renzini}, {Ho}, {Mannucci},
  {Daddi}, {Gao}, {Maiolino}, {Zhang}, {Gu}, {Li}, {Lilly}, \&
  {Yuan}}]{2021Dou}
{Dou}, J., {Peng}, Y., {Renzini}, A., {et~al.} 2021, \apj, 907, 114

\bibitem[{{Eales} {et~al.}(2020){Eales}, {Eales}, \& {de
  Vis}}]{2020MNRAS.491...69E}
{Eales}, S., {Eales}, O., \& {de Vis}, P. 2020, \mnras, 491, 69

\bibitem[{{Ellison} {et~al.}(2020{\natexlab{a}}){Ellison}, {Thorp}, {Pan},
  {Lin}, {Scudder}, {Bluck}, {S{\'a}nchez}, \& {Sargent}}]{2020MNRAS.492.6027E}
{Ellison}, S.~L., {Thorp}, M.~D., {Pan}, H.-A., {et~al.} 2020{\natexlab{a}},
  \mnras, 492, 6027

\bibitem[{{Ellison} {et~al.}(2020{\natexlab{b}}){Ellison}, {Thorp}, {Lin},
  {Pan}, {Bluck}, {Scudder}, {Teimoorinia}, {S{\'a}nchez}, \&
  {Sargent}}]{2020MNRAS.493L..39E}
{Ellison}, S.~L., {Thorp}, M.~D., {Lin}, L., {et~al.} 2020{\natexlab{b}},
  \mnras, 493, L39

\bibitem[{{Elmegreen} \& {Efremov}(1997)}]{1997ApJ...480..235E}
{Elmegreen}, B.~G., \& {Efremov}, Y.~N. 1997, \apj, 480, 235

\bibitem[{{F{\"o}rster Schreiber} {et~al.}(2019){F{\"o}rster Schreiber},
  {{\"U}bler}, {Davies}, {Genzel}, {Wisnioski}, {Belli}, {Shimizu}, {Lutz},
  {Fossati}, {Herrera-Camus}, {Mendel}, {Tacconi}, {Wilman}, {Beifiori},
  {Brammer}, {Burkert}, {Carollo}, {Davies}, {Eisenhauer}, {Fabricius},
  {Lilly}, {Momcheva}, {Naab}, {Nelson}, {Price}, {Renzini}, {Saglia},
  {Sternberg}, {van Dokkum}, \& {Wuyts}}]{2019ApJ...875...21F}
{F{\"o}rster Schreiber}, N.~M., {{\"U}bler}, H., {Davies}, R.~L., {et~al.}
  2019, \apj, 875, 21

\bibitem[{{Fossati} {et~al.}(2017){Fossati}, {Wilman}, {Mendel}, {Saglia},
  {Galametz}, {Beifiori}, {Bender}, {Chan}, {Fabricius}, {Bandara}, {Brammer},
  {Davies}, {F{\"o}rster Schreiber}, {Genzel}, {Hartley}, {Kulkarni}, {Lang},
  {Momcheva}, {Nelson}, {Skelton}, {Tacconi}, {Tadaki}, {{\"U}bler}, {van
  Dokkum}, {Wisnioski}, {Whitaker}, {Wuyts}, \& {Wuyts}}]{2017ApJ...835..153F}
{Fossati}, M., {Wilman}, D.~J., {Mendel}, J.~T., {et~al.} 2017, \apj, 835, 153

\bibitem[{{Gensior} {et~al.}(2020){Gensior}, {Kruijssen}, \&
  {Keller}}]{2020MNRAS.495..199G}
{Gensior}, J., {Kruijssen}, J.~M.~D., \& {Keller}, B.~W. 2020, \mnras, 495, 199

\bibitem[{{Genzel} {et~al.}(2010){Genzel}, {Tacconi}, {Gracia-Carpio},
  {Sternberg}, {Cooper}, {Shapiro}, {Bolatto}, {Bouch{\'e}}, {Bournaud},
  {Burkert}, {Combes}, {Comerford}, {Cox}, {Davis}, {Schreiber},
  {Garcia-Burillo}, {Lutz}, {Naab}, {Neri}, {Omont}, {Shapley}, \&
  {Weiner}}]{2010Genzel}
{Genzel}, R., {Tacconi}, L.~J., {Gracia-Carpio}, J., {et~al.} 2010, \mnras,
  407, 2091

\bibitem[{{Genzel} {et~al.}(2012){Genzel}, {Tacconi}, {Combes}, {Bolatto},
  {Neri}, {Sternberg}, {Cooper}, {Bouch{\'e}}, {Bournaud}, {Burkert},
  {Comerford}, {Cox}, {Davis}, {F{\"o}rster Schreiber}, {Garcia-Burillo},
  {Gracia-Carpio}, {Lutz}, {Naab}, {Newman}, {Saintonge}, {Shapiro}, {Shapley},
  \& {Weiner}}]{2012ApJ...746...69G}
{Genzel}, R., {Tacconi}, L.~J., {Combes}, F., {et~al.} 2012, \apj, 746, 69

\bibitem[{{Genzel} {et~al.}(2014){Genzel}, {F{\"o}rster Schreiber}, {Lang},
  {Tacchella}, {Tacconi}, {Wuyts}, {Bandara}, {Burkert}, {Buschkamp},
  {Carollo}, {Cresci}, {Davies}, {Eisenhauer}, {Hicks}, {Kurk}, {Lilly},
  {Lutz}, {Mancini}, {Naab}, {Newman}, {Peng}, {Renzini}, {Shapiro Griffin},
  {Sternberg}, {Vergani}, {Wisnioski}, {Wuyts}, \&
  {Zamorani}}]{2014ApJ...785...75G}
{Genzel}, R., {F{\"o}rster Schreiber}, N.~M., {Lang}, P., {et~al.} 2014, \apj,
  785, 75

\bibitem[{{Genzel} {et~al.}(2015){Genzel}, {Tacconi}, {Lutz}, {Saintonge},
  {Berta}, {Magnelli}, {Combes}, {Garc{\'\i}a-Burillo}, {Neri}, {Bolatto},
  {Contini}, {Lilly}, {Boissier}, {Boone}, {Bouch{\'e}}, {Bournaud}, {Burkert},
  {Carollo}, {Colina}, {Cooper}, {Cox}, {Feruglio}, {F{\"o}rster Schreiber},
  {Freundlich}, {Gracia-Carpio}, {Juneau}, {Kovac}, {Lippa}, {Naab}, {Salome},
  {Renzini}, {Sternberg}, {Walter}, {Weiner}, {Weiss}, \&
  {Wuyts}}]{Genzel:2015fq}
{Genzel}, R., {Tacconi}, L.~J., {Lutz}, D., {et~al.} 2015, \apj, 800, 20

\bibitem[{{Giovanelli} \& {Haynes}(1985)}]{1985ApJ...292..404G}
{Giovanelli}, R., \& {Haynes}, M.~P. 1985, \apj, 292, 404

\bibitem[{{Goldsmith} {et~al.}(2007){Goldsmith}, {Li}, \&
  {Kr{\v{c}}o}}]{2007ApJ...654..273G}
{Goldsmith}, P.~F., {Li}, D., \& {Kr{\v{c}}o}, M. 2007, \apj, 654, 273

\bibitem[{{Granato} {et~al.}(2004){Granato}, {De Zotti}, {Silva}, {Bressan}, \&
  {Danese}}]{2004ApJ...600..580G}
{Granato}, G.~L., {De Zotti}, G., {Silva}, L., {Bressan}, A., \& {Danese}, L.
  2004, \apj, 600, 580

\bibitem[{{Gunn} \& {Gott}(1972)}]{1972ApJ...176....1G}
{Gunn}, J.~E., \& {Gott}, J.~Richard, I. 1972, \apj, 176, 1

\bibitem[{{Guo} {et~al.}(2017){Guo}, {Bell}, {Lu}, {Koo}, {Faber}, {Koekemoer},
  {Kurczynski}, {Lee}, {Papovich}, {Chen}, {Dekel}, {Ferguson}, {Fontana},
  {Giavalisco}, {Kocevski}, {Nayyeri}, {P{\'e}rez-Gonz{\'a}lez}, {Pforr},
  {Rodr{\'\i}guez-Puebla}, \& {Santini}}]{2017ApJ...841L..22G}
{Guo}, Y., {Bell}, E.~F., {Lu}, Y., {et~al.} 2017, \apjl, 841, L22

\bibitem[{{Haynes} {et~al.}(2011){Haynes}, {Giovanelli}, {Martin}, {Hess},
  {Saintonge}, {Adams}, {Hallenbeck}, {Hoffman}, {Huang}, {Kent}, {Koopmann},
  {Papastergis}, {Stierwalt}, {Balonek}, {Craig}, {Higdon}, {Kornreich},
  {Miller}, {O'Donoghue}, {Olowin}, {Rosenberg}, {Spekkens}, {Troischt}, \&
  {Wilcots}}]{Haynes:2011en}
{Haynes}, M.~P., {Giovanelli}, R., {Martin}, A.~M., {et~al.} 2011, \aj, 142,
  170

\bibitem[{{Huang} \& {Kauffmann}(2014)}]{Huang:2014ko}
{Huang}, M.-L., \& {Kauffmann}, G. 2014, \mnras, 443, 1329

\bibitem[{{Ilbert} {et~al.}(2015){Ilbert}, {Arnouts}, {Le Floc'h}, {Aussel},
  {Bethermin}, {Capak}, {Hsieh}, {Kajisawa}, {Karim}, {Le F{\`e}vre}, {Lee},
  {Lilly}, {McCracken}, {Michel-Dansac}, {Moutard}, {Renzini}, {Salvato},
  {Sanders}, {Scoville}, {Sheth}, {Silverman}, {Smol{\v{c}}i{\'c}},
  {Taniguchi}, \& {Tresse}}]{2015A&A...579A...2I}
{Ilbert}, O., {Arnouts}, S., {Le Floc'h}, E., {et~al.} 2015, \aap, 579, A2

\bibitem[{{Janowiecki} {et~al.}(2017){Janowiecki}, {Catinella}, {Cortese},
  {Saintonge}, {Brown}, \& {Wang}}]{2017Janowiecki}
{Janowiecki}, S., {Catinella}, B., {Cortese}, L., {et~al.} 2017, \mnras, 466,
  4795

\bibitem[{{Kauffmann} {et~al.}(2004){Kauffmann}, {White}, {Heckman},
  {M{\'e}nard}, {Brinchmann}, {Charlot}, {Tremonti}, \&
  {Brinkmann}}]{2004MNRAS.353..713K}
{Kauffmann}, G., {White}, S. D.~M., {Heckman}, T.~M., {et~al.} 2004, \mnras,
  353, 713

\bibitem[{{Kennicutt}(1998)}]{1998Kennicutt}
{Kennicutt}, Robert~C., J. 1998, \apj, 498, 541

\bibitem[{{Krumholz} {et~al.}(2018){Krumholz}, {Burkhart}, {Forbes}, \&
  {Crocker}}]{2018MNRAS.477.2716K}
{Krumholz}, M.~R., {Burkhart}, B., {Forbes}, J.~C., \& {Crocker}, R.~M. 2018,
  \mnras, 477, 2716

\bibitem[{{Krumholz} {et~al.}(2009){Krumholz}, {McKee}, \&
  {Tumlinson}}]{2009Krumholz}
{Krumholz}, M.~R., {McKee}, C.~F., \& {Tumlinson}, J. 2009, \apj, 693, 216

\bibitem[{{Larson} {et~al.}(1980){Larson}, {Tinsley}, \&
  {Caldwell}}]{1980ApJ...237..692L}
{Larson}, R.~B., {Tinsley}, B.~M., \& {Caldwell}, C.~N. 1980, \apj, 237, 692

\bibitem[{{Lequeux} {et~al.}(1979){Lequeux}, {Peimbert}, {Rayo}, {Serrano}, \&
  {Torres-Peimbert}}]{1979A&A....80..155L}
{Lequeux}, J., {Peimbert}, M., {Rayo}, J.~F., {Serrano}, A., \&
  {Torres-Peimbert}, S. 1979, \aap, 500, 145

\bibitem[{{Lilly} {et~al.}(2013){Lilly}, {Carollo}, {Pipino}, {Renzini}, \&
  {Peng}}]{Lilly:2013ko}
{Lilly}, S.~J., {Carollo}, C.~M., {Pipino}, A., {Renzini}, A., \& {Peng}, Y.
  2013, \apj, 772, 119

\bibitem[{{Liszt}(2007)}]{2007A&A...461..205L}
{Liszt}, H.~S. 2007, \aap, 461, 205

\bibitem[{{Madau} \& {Dickinson}(2014)}]{2014ARA&A..52..415M}
{Madau}, P., \& {Dickinson}, M. 2014, \araa, 52, 415

\bibitem[{{Maiolino} \& {Mannucci}(2019)}]{2019A&ARv..27....3M}
{Maiolino}, R., \& {Mannucci}, F. 2019, \aapr, 27, 3

\bibitem[{{Mannucci} {et~al.}(2010){Mannucci}, {Cresci}, {Maiolino}, {Marconi},
  \& {Gnerucci}}]{2010MNRAS.408.2115M}
{Mannucci}, F., {Cresci}, G., {Maiolino}, R., {Marconi}, A., \& {Gnerucci}, A.
  2010, \mnras, 408, 2115

\bibitem[{{Martig} {et~al.}(2009){Martig}, {Bournaud}, {Teyssier}, \&
  {Dekel}}]{2009ApJ...707..250M}
{Martig}, M., {Bournaud}, F., {Teyssier}, R., \& {Dekel}, A. 2009, \apj, 707,
  250

\bibitem[{{Martin} {et~al.}(2005){Martin}, {Fanson}, {Schiminovich},
  {Morrissey}, {Friedman}, {Barlow}, {Conrow}, {Grange}, {Jelinsky},
  {Milliard}, {Siegmund}, {Bianchi}, {Byun}, {Donas}, {Forster}, {Heckman},
  {Lee}, {Madore}, {Malina}, {Neff}, {Rich}, {Small}, {Surber}, {Szalay},
  {Welsh}, \& {Wyder}}]{Martin:2005}
{Martin}, D.~C., {Fanson}, J., {Schiminovich}, D., {et~al.} 2005, \apjl, 619,
  L1

\bibitem[{{McKee} \& {Krumholz}(2010)}]{2010McKee}
{McKee}, C.~F., \& {Krumholz}, M.~R. 2010, \apj, 709, 308

\bibitem[{{Mendel} {et~al.}(2014){Mendel}, {Simard}, {Palmer}, {Ellison}, \&
  {Patton}}]{2014ApJS..210....3M}
{Mendel}, J.~T., {Simard}, L., {Palmer}, M., {Ellison}, S.~L., \& {Patton},
  D.~R. 2014, \apjs, 210, 3

\bibitem[{{Morselli} {et~al.}(2020){Morselli}, {Rodighiero}, {Enia},
  {Corbelli}, {Casasola}, {Rodr{\'\i}guez-Mu{\~n}oz}, {Renzini}, {Tacchella},
  {Baronchelli}, {Bianchi}, {Cassata}, {Franceschini}, {Mancini}, {Negrello},
  {Popesso}, \& {Romano}}]{2020MNRAS.496.4606M}
{Morselli}, L., {Rodighiero}, G., {Enia}, A., {et~al.} 2020, \mnras, 496, 4606

\bibitem[{{Morselli} {et~al.}(2021){Morselli}, {Renzini}, {Enia}, \&
  {Rodighiero}}]{2021Morselli}
{Morselli}, L., {Renzini}, A., {Enia}, A., \& {Rodighiero}, G. 2021, \mnras,
  502, L85

\bibitem[{{Pan} {et~al.}(2019){Pan}, {Peng}, {Zheng}, {Wang}, \&
  {Kong}}]{2019ApJ...885L..14P}
{Pan}, Z., {Peng}, Y., {Zheng}, X., {Wang}, J., \& {Kong}, X. 2019, \apjl, 885,
  L14

\bibitem[{{Peng} {et~al.}(2015){Peng}, {Maiolino}, \& {Cochrane}}]{Peng:2015bq}
{Peng}, Y., {Maiolino}, R., \& {Cochrane}, R. 2015, \nat, 521, 192

\bibitem[{{Peng} \& {Maiolino}(2014)}]{Peng:2014hn}
{Peng}, Y.-j., \& {Maiolino}, R. 2014, \mnras, 443, 3643

\bibitem[{{Peng} \& {Renzini}(2020)}]{2020MNRAS.491L..51P}
{Peng}, Y.-j., \& {Renzini}, A. 2020, \mnras, 491, L51

\bibitem[{{Peng} {et~al.}(2010){Peng}, {Lilly}, {Kova{\v{c}}}, {Bolzonella},
  {Pozzetti}, {Renzini}, {Zamorani}, {Ilbert}, {Knobel}, {Iovino}, {Maier},
  {Cucciati}, {Tasca}, {Carollo}, {Silverman}, {Kampczyk}, {de Ravel},
  {Sanders}, {Scoville}, {Contini}, {Mainieri}, {Scodeggio}, {Kneib}, {Le
  F{\`e}vre}, {Bardelli}, {Bongiorno}, {Caputi}, {Coppa}, {de la Torre},
  {Franzetti}, {Garilli}, {Lamareille}, {Le Borgne}, {Le Brun}, {Mignoli},
  {Perez Montero}, {Pello}, {Ricciardelli}, {Tanaka}, {Tresse}, {Vergani},
  {Welikala}, {Zucca}, {Oesch}, {Abbas}, {Barnes}, {Bordoloi}, {Bottini},
  {Cappi}, {Cassata}, {Cimatti}, {Fumana}, {Hasinger}, {Koekemoer},
  {Leauthaud}, {Maccagni}, {Marinoni}, {McCracken}, {Memeo}, {Meneux}, {Nair},
  {Porciani}, {Presotto}, \& {Scaramella}}]{Peng:2010gn}
{Peng}, Y.-j., {Lilly}, S.~J., {Kova{\v{c}}}, K., {et~al.} 2010, \apj, 721, 193

\bibitem[{{Quilis} {et~al.}(2000){Quilis}, {Moore}, \&
  {Bower}}]{2000Sci...288.1617Q}
{Quilis}, V., {Moore}, B., \& {Bower}, R. 2000, Science, 288, 1617

\bibitem[{{Renzini}(2020)}]{2020MNRAS.495L..42R}
{Renzini}, A. 2020, \mnras, 495, L42


\bibitem[{{Saintonge} {et~al.}(2011){Saintonge}, {Kauffmann}, {Wang}, {Kramer},
  {Tacconi}, {Buchbender}, {Catinella}, {Graci{\'a}-Carpio}, {Cortese},
  {Fabello}, {Fu}, {Genzel}, {Giovanelli}, {Guo}, {Haynes}, {Heckman},
  {Krumholz}, {Lemonias}, {Li}, {Moran}, {Rodriguez-Fernand ez},
  {Schiminovich}, {Schuster}, \& {Sievers}}]{Saintonge:2011ey}
{Saintonge}, A., {Kauffmann}, G., {Wang}, J., {et~al.} 2011, \mnras, 415, 61

\bibitem[{{Saintonge} {et~al.}(2012){Saintonge}, {Tacconi}, {Fabello}, {Wang},
  {Catinella}, {Genzel}, {Graci{\'a}-Carpio}, {Kramer}, {Moran}, {Heckman},
  {Schiminovich}, {Schuster}, \& {Wuyts}}]{2012Saintonge}
{Saintonge}, A., {Tacconi}, L.~J., {Fabello}, S., {et~al.} 2012, \apj, 758, 73

\bibitem[{{Saintonge} {et~al.}(2016){Saintonge}, {Catinella}, {Cortese},
  {Genzel}, {Giovanelli}, {Haynes}, {Janowiecki}, {Kramer}, {Lutz},
  {Schiminovich}, {Tacconi}, {Wuyts}, \& {Accurso}}]{2016MNRAS.462.1749S}
{Saintonge}, A., {Catinella}, B., {Cortese}, L., {et~al.} 2016, \mnras, 462,
  1749

\bibitem[{{Saintonge} {et~al.}(2017){Saintonge}, {Catinella}, {Tacconi},
  {Kauffmann}, {Genzel}, {Cortese}, {Dav{\'e}}, {Fletcher},
  {Graci{\'a}-Carpio}, {Kramer}, {Heckman}, {Janowiecki}, {Lutz}, {Rosario},
  {Schiminovich}, {Schuster}, {Wang}, {Wuyts}, {Borthakur}, {Lamperti}, \&
  {Roberts-Borsani}}]{2017Saintonge}
{Saintonge}, A., {Catinella}, B., {Tacconi}, L.~J., {et~al.} 2017, \apjs, 233,
  22

\bibitem[{{Salim} {et~al.}(2007){Salim}, {Rich}, {Charlot}, {Brinchmann},
  {Johnson}, {Schiminovich}, {Seibert}, {Mallery}, {Heckman}, {Forster},
  {Friedman}, {Martin}, {Morrissey}, {Neff}, {Small}, {Wyder}, {Bianchi},
  {Donas}, {Lee}, {Madore}, {Milliard}, {Szalay}, {Welsh}, \& {Yi}}]{2007Salim}
{Salim}, S., {Rich}, R.~M., {Charlot}, S., {et~al.} 2007, \apjs, 173, 267


\bibitem[{{Sargent} {et~al.}(2014){Sargent}, {Daddi}, {B{\'e}thermin},
  {Aussel}, {Magdis}, {Hwang}, {Juneau}, {Elbaz}, \& {da
  Cunha}}]{Sargent:2014ky}
{Sargent}, M.~T., {Daddi}, E., {B{\'e}thermin}, M., {et~al.} 2014, \apj, 793,
  19

\bibitem[{{Scoville} {et~al.}(2017){Scoville}, {Lee}, {Vanden Bout},
  {Diaz-Santos}, {Sanders}, {Darvish}, {Bongiorno}, {Casey}, {Murchikova},
  {Koda}, {Capak}, {Vlahakis}, {Ilbert}, {Sheth}, {Morokuma-Matsui}, {Ivison},
  {Aussel}, {Laigle}, {McCracken}, {Armus}, {Pope}, {Toft}, \&
  {Masters}}]{2017ApJ...837..150S}
{Scoville}, N., {Lee}, N., {Vanden Bout}, P., {et~al.} 2017, \apj, 837, 150

\bibitem[{{Silk}(1997)}]{1997Silk}
{Silk}, J. 1997, \apj, 481, 703

\bibitem[{{Simard} {et~al.}(2011){Simard}, {Mendel}, {Patton}, {Ellison}, \&
  {McConnachie}}]{Simard2011}
{Simard}, L., {Mendel}, J.~T., {Patton}, D.~R., {Ellison}, S.~L., \&
  {McConnachie}, A.~W. 2011, \apjs, 196, 11

\bibitem[{{Somerville} \& {Dav{\'e}}(2015)}]{2015ARA&A..53...51S}
{Somerville}, R.~S., \& {Dav{\'e}}, R. 2015, \araa, 53, 51

\bibitem[{{Song} {et~al.}(2020){Song}, {Laigle}, {Hwang}, {Devriendt},
  {Dubois}, {Kraljic}, {Pichon}, {Slyz}, \& {Smith}}]{2020arXiv200900013S}
{Song}, H., {Laigle}, C., {Hwang}, H.~S., {et~al.} 2020, arXiv e-prints,
  arXiv:2009.00013

\bibitem[{{Speagle} {et~al.}(2014){Speagle}, {Steinhardt}, {Capak}, \&
  {Silverman}}]{Speagle:2014dd}
{Speagle}, J.~S., {Steinhardt}, C.~L., {Capak}, P.~L., \& {Silverman}, J.~D.
  2014, \apjs, 214, 15

\bibitem[{{Springob} {et~al.}(2005){Springob}, {Haynes}, {Giovanelli}, \&
  {Kent}}]{Springob:2005vt}
{Springob}, C.~M., {Haynes}, M.~P., {Giovanelli}, R., \& {Kent}, B.~R. 2005,
  \apjs, 160, 149


\bibitem[{{Tacconi} {et~al.}(2020){Tacconi}, {Genzel}, \&
  {Sternberg}}]{2020Tacconi}
{Tacconi}, L.~J., {Genzel}, R., \& {Sternberg}, A. 2020, \araa, 58, 157

\bibitem[{{Tacconi} {et~al.}(2018){Tacconi}, {Genzel}, {Saintonge}, {Combes},
  {Garc{\'\i}a-Burillo}, {Neri}, {Bolatto}, {Contini}, {F{\"o}rster Schreiber},
  {Lilly}, {Lutz}, {Wuyts}, {Accurso}, {Boissier}, {Boone}, {Bouch{\'e}},
  {Bournaud}, {Burkert}, {Carollo}, {Cooper}, {Cox}, {Feruglio}, {Freundlich},
  {Herrera-Camus}, {Juneau}, {Lippa}, {Naab}, {Renzini}, {Salome}, {Sternberg},
  {Tadaki}, {{\"U}bler}, {Walter}, {Weiner}, \& {Weiss}}]{Tacconi:2018jb}
{Tacconi}, L.~J., {Genzel}, R., {Saintonge}, A., {et~al.} 2018, \apj, 853, 179

\bibitem[{{Tremonti} {et~al.}(2004){Tremonti}, {Heckman}, {Kauffmann},
  {Brinchmann}, {Charlot}, {White}, {Seibert}, {Peng}, {Schlegel}, {Uomoto},
  {Fukugita}, \& {Brinkmann}}]{2004ApJ...613..898T}
{Tremonti}, C.~A., {Heckman}, T.~M., {Kauffmann}, G., {et~al.} 2004, \apj, 613,
  898

\bibitem[{{Trussler} {et~al.}(2020{\natexlab{a}}){Trussler}, {Maiolino},
  {Maraston}, {Peng}, {Thomas}, {Goddard}, \& {Lian}}]{2020MNRAS.491.5406T}
{Trussler}, J., {Maiolino}, R., {Maraston}, C., {et~al.} 2020{\natexlab{a}},
  \mnras, 491, 5406

\bibitem[{{Trussler} {et~al.}(2020{\natexlab{b}}){Trussler}, {Maiolino},
  {Maraston}, {Peng}, {Thomas}, {Goddard}, \& {Lian}}]{2020MNRAS.tmp.3344T}
{Trussler}, J., {Maiolino}, R., {Maraston}, C., {et~al.} 2020{\natexlab{b}}, \mnras, doi:10.1093/mnras/staa3545

\bibitem[{{van de Voort} {et~al.}(2011){van de Voort}, {Schaye}, {Booth}, \&
  {Dalla Vecchia}}]{2011MNRAS.415.2782V}
{van de Voort}, F., {Schaye}, J., {Booth}, C.~M., \& {Dalla Vecchia}, C. 2011,
  \mnras, 415, 2782

\bibitem[{{van Driel} {et~al.}(2001){van Driel}, {Gao}, \&
  {Monnier-Ragaigne}}]{2001van}
{van Driel}, W., {Gao}, Y., \& {Monnier-Ragaigne}, D. 2001, \aap, 368, 64

\bibitem[{{Veilleux} {et~al.}(2020){Veilleux}, {Maiolino}, {Bolatto}, \&
  {Aalto}}]{2020Veilleux}
{Veilleux}, S., {Maiolino}, R., {Bolatto}, A.~D., \& {Aalto}, S. 2020, \aapr,
  28, 2

\bibitem[{{Walter} {et~al.}(2020){Walter}, {Carilli}, {Neeleman}, {Decarli},
  {Popping}, {Somerville}, {Aravena}, {Bertoldi}, {Boogaard}, {Cox}, {da
  Cunha}, {Magnelli}, {Obreschkow}, {Riechers}, {Rix}, {Smail}, {Weiss},
  {Assef}, {Bauer}, {Bouwens}, {Contini}, {Cortes}, {Daddi}, {Diaz-Santos},
  {Gonz{\'a}lez-L{\'o}pez}, {Hennawi}, {Hodge}, {Inami}, {Ivison}, {Oesch},
  {Sargent}, {van der Werf}, {Wagg}, \& {Yung}}]{2020Walter}
{Walter}, F., {Carilli}, C., {Neeleman}, M., {et~al.} 2020, \apj, 902, 111

\bibitem[{{White} \& {Rees}(1978)}]{1978MNRAS.183..341W}
{White}, S.~D.~M., \& {Rees}, M.~J. 1978, \mnras, 183, 341

\bibitem[{{Wong} \& {Blitz}(2002)}]{2002Wong}
{Wong}, T., \& {Blitz}, L. 2002, \apj, 569, 157

\bibitem[{{Yang} {et~al.}(2007){Yang}, {Mo}, {van den Bosch}, {Pasquali}, {Li},
  \& {Barden}}]{Yang:2007}
{Yang}, X., {Mo}, H.~J., {van den Bosch}, F.~C., {et~al.} 2007, \apj, 671, 153

\bibitem[{{Zhang} {et~al.}(2019){Zhang}, {Peng}, {Ho}, {Maiolino}, {Dekel},
  {Guo}, {Mannucci}, {Li}, {Yuan}, {Renzini}, {Dou}, {Guo}, {Man}, \&
  {Li}}]{2019ApJ...884L..52Z}
{Zhang}, C., {Peng}, Y., {Ho}, L.~C., {et~al.} 2019, \apjl, 884, L52

\bibitem[{{Zhang} {et~al.}(2021){Zhang}, {Peng}, {Ho}, {Maiolino}, {Renzini},
  {Mannucci}, {Dekel}, {Guo}, {Li}, {Yuan}, {Lilly}, {Dou}, {Guo}, {Man}, {Li},
  \& {Shi}}]{2021zhang}
{Zhang}, C., {Peng}, Y., {Ho}, L.~C., {et~al.} 2021, arXiv e-prints, arXiv:2104.07045

\bibitem[{{Zinger} {et~al.}(2020){Zinger}, {Pillepich}, {Nelson}, {Weinberger},
  {Pakmor}, {Springel}, {Hernquist}, {Marinacci}, \&
  {Vogelsberger}}]{2020Zinger}
{Zinger}, E., {Pillepich}, A., {Nelson}, D., {et~al.} 2020, \mnras, 499, 768

\end{thebibliography}
\end{document}